\documentclass[letterpaper,12pt]{article}
\usepackage{array}
\usepackage{dcolumn}
\usepackage{float}
\usepackage{natbib}
\usepackage{graphicx}
\usepackage{color}
\usepackage{amsmath}
\usepackage{amsfonts}
\usepackage{amsopn}
\usepackage{amsthm}
\usepackage{lscape}
\usepackage{multirow}
\newtheorem{thm}{Theorem}

\theoremstyle{definition}

\theoremstyle{remark}
\newtheorem{rmk}{Remark}
\newtheorem*{rmk.nonumber}{Remark}
\DeclareMathOperator{\pr}{P}
\DeclareMathOperator{\epn}{E}
\DeclareMathOperator{\var}{var}

\DeclareMathOperator{\logit}{logit}

\newcommand{\bmW}{\boldsymbol W}%
\newcommand{\bmO}{\boldsymbol O}%
\newcommand{\bmD}{\boldsymbol D}%
\newcommand{\bmX}{\boldsymbol X}%
\newcommand{\bmc}{\boldsymbol c}%
\newcommand{\bmw}{\boldsymbol w}%
\newcommand{\bmeta}{\mbox{\boldmath${\eta}$}}%
\newcommand{\bmgamma}{\mbox{\boldmath${\gamma}$}}%
\newcommand{\bmalpha}{\mbox{\boldmath${\alpha}$}}%

\renewcommand{\baselinestretch}{1.8}

\begin{document}
\begin{center}
{\LARGE An Adaptive Design for Optimizing Treatment Assignment in Randomized Clinical Trials}
\end{center}
\begin{center}
Wei Zhang$^{1}$, Zhiwei Zhang$^{2,*}$, and Aiyi Liu$^{3}$\\
$^1$State Key Laboratory of Mathematical Sciences,
Academy of Mathematics and Systems Science, Chinese Academy of Sciences, Beijing, China\\
$^2$Biostatistics Innovation Group, Gilead Sciences, Foster City, California, USA\\
$^3$Biostatistics and Bioinformatics Branch, Division of Population Health Research, \textit{Eunice Kennedy Shriver} National Institute of Child Health and Human Development, National Institutes of Health, Bethesda, Maryland, USA\\
$^*$Zhiwei.Zhang6@gilead.com
\end{center}

\vspace{0.5cm}
\centerline{\bf Abstract}
The treatment assignment mechanism in a randomized clinical trial can be optimized for statistical efficiency within a specified class of randomization mechanisms. Optimal designs of this type have been characterized in terms of the variances of potential outcomes conditional on baseline covariates. Approximating these optimal designs requires information about the conditional variance functions, which is often unavailable or unreliable at the design stage. As a practical solution to this dilemma, we propose a multi-stage adaptive design that allows the treatment assignment mechanism to be modified at interim analyses based on accruing information about the conditional variance functions. This adaptation has profound implications on the distribution of trial data, which need to be accounted for in treatment effect estimation. We consider a class of treatment effect estimators that are consistent and asymptotically normal, identify the most efficient estimator within this class, and approximate the most efficient estimator by substituting estimates of unknown quantities. Simulation results indicate that, when there is little or no prior information available, the proposed design can bring substantial efficiency gains over conventional one-stage designs based on the same prior information. The methodology is illustrated with real data from a completed trial in stroke.

\vspace{.5cm}

\newpage

\section{Introduction}\label{intro}

Randomized clinical trials (RCTs) are widely considered the gold standard for comparing an experimental treatment with a control treatment. RCTs can be highly expensive in terms of time and money, making it crucial to optimize clinical trial design and analysis for statistical efficiency. The analysis of trial data can be made more efficient by incorporating information from baseline covariates that are related to clinical outcomes, a practice commonly known as covariate adjustment. There are many statistical methods for covariate adjustment \citep[e.g.,][]{t08,z08,m09,w19,zm19,y23,b25}, which provide covariate-adjusted treatment effect estimators that are asymptotically unbiased and generally more efficient than unadjusted estimators. 
Covariate adjustment has drawn a great deal of attention and serious consideration from practitioners and regulators \citep{fda23}.

Treatment assignment is a key aspect of trial design that can be optimized for statistical efficiency. For unadjusted treatment effect estimators based on treatment-specific sample means, the optimal allocation, known as the Neyman allocation \citep{n34}, assigns group sizes in proportion to the standard deviation of the primary outcome within each treatment group. Such estimators are generally inefficient as they do not make use of baseline covariate information. \citet{z23} derive optimal allocation ratios for covariate-adjusted treatment effect estimators, including the optimal one that attains the nonparametric information bound. The optimal allocation ratio for the optimal estimator can be regarded as an optimal design that maximizes the nonparametric information bound. In addition to conventional covariate-independent randomization (CIR), \citet{z23} also consider covariate-dependent randomization (CDR), which resembles observational studies except that the propensity score is specified and known by the investigator. The propensity score in CDR can be optimized for (locally) efficient treatment effect estimators available from the causal inference literature \citep[e.g.,][]{vr03,vr11}. The optimal propensity score for the efficient estimator represents the optimal CDR design that maximizes the nonparametric information bound over all CDR designs (including CIR designs as special cases). The optimal CIR and CDR designs differ fundamentally from covariate-adaptive randomization designs \citep[e.g.,][]{rs08}, which improve efficiency for individual estimators but do not change the nonparametric information bound relative to simple randomization \citep{r23}.

The optimal CIR and CDR designs are characterized using the variances of potential outcomes conditional on baseline covariates. Approximating these optimal designs requires preliminary information on the conditional variance functions. In practice, preliminary estimates of the conditional variance functions are often unavailable or unreliable, making it difficult to construct efficient designs based on the optimality results of \citet{z23}. Their simulation results indicate that unreliable estimates of conditional variance functions can result in ill-optimized designs that perform poorly, especially when the optimal CDR design is targeted.

As a concrete example, consider the National Institute of Neurological Disorders and Stroke (NINDS) trial of recombinant tissue plasminogen activator (rt-PA) for treating acute ischemic stroke \citep{n95}. In this trial, a total of 624 eligible patients were randomized 1:1 to receive rt-PA or placebo within 3 hours of stroke onset. Patients were evaluated at baseline (before randomization) and at 24 hours, 3 months, and 1 year post-treatment. The primary efficacy analysis found statistically significant improvements in clinical outcomes at 3 months. \citet{z23} show that, in this setting, optimal designs based on three dichotomous baseline covariates can reduce the variance by as much as 25\% in treatment effect estimation. However, this improvement is based on the assumption that the true conditional variance functions are known at the design stage. This assumption is unlikely to hold in practice, which begs the question: what can be done to improve design efficiency (in terms of treatment assignment) with little or no information available about the conditional variance functions?

In this article, we propose an adaptive design as a practical answer to the above question. The proposed design consists of multiple stages separated by one or more interim analyses. In the first stage, treatment assignment may follow the standard $1:1$ CIR design (if no prior information is available) or a cautiously optimized CIR or CDR design based on prior information. At each interim analysis, the conditional variance functions are (re-)estimated with updated data, and the updated estimates are used to optimize treatment assignment in the next stage. In general, as information accrues over time, treatment assignment optimization can become more ambitious in terms of the set of covariates to consider, the amount of coarsening to apply (to the covariates), and the class of designs (CIR versus CDR) to optimize over. 

While conceptually simple and intuitive, the proposed adaptive design does present serious analytical challenges in treatment effect estimation and inference. Because the treatment assignment mechanism may change over time, individual patient data are generally not identically distributed between different stages. Because the treatment assignment mechanism in a later stage may depend on data from previous stages, independence between patients is no longer guaranteed except within the first stage. We consider a class of treatment effect estimators indexed by a set of weights (for combining data across stages) and a set of augmentation functions (for incorporating covariate information). Each estimator in this class is consistent and asymptotically normal, and its asymptotic variance depends on the weights and augmentation functions. The optimal weights and augmentation functions, which together minimize the asymptotic variance of the estimator, are characterized and straightforward to estimate. When estimated weights and augmentation functions are substituted, the resulting treatment effect estimator remains consistent and asymptotically normal, and is asymptotically equivalent to the one based on the limiting weights and augmentation functions.


\section{Preliminaries}\label{pre}

We start by formulating the estimation problem using potential outcomes. For a generic patient in the study population, let $Y(a)$ denote the potential outcome for treatment $a$, where $a=1$ for an experimental treatment and $a=0$ for a control treatment, which may be placebo or standard of care. For each $a\in\{0,1\}$, we assume $\epn\{Y(a)^2\}<\infty$ and write $\mu_a=\epn\{Y(a)\}$. Suppose the treatment effect of interest is $\delta=g(\mu_1)-g(\mu_0)$, where $g$ is a smooth and strictly increasing function specified by the investigator. Popular choices for $g$ include identity (for continuous and other types of outcomes), logarithm (for outcomes with positive means), and logit (for binary outcomes). Let $\bmW$ be a vector of available baseline covariates (i.e., pre-treatment patient characteristics) that may be associated with either or both of $Y(1)$ and $Y(0)$.

An RCT takes a random sample of patients from the study population, assigns a treatment $A\in\{0,1\}$ to each participant in a randomized fashion, and measures the resulting outcome $Y=Y(A)=AY(1)+(1-A)Y(0)$. In current RCT practice, treatment assignment usually follows CIR, where $A$ is independent of all baseline variables. In particular, we have
$$
\pr\{A=1|\bmW,Y(1),Y(0)\}=\pr(A=1)=\pi\in(0,1),
$$
where $\pi$ is a pre-specified constant. In the CDR design proposed by \citet{z23}, $A$ is allowed to depend on observed baseline covariates but not on potential outcomes.  Specifically,
$$
\pr\{A=1|\bmW,Y(1),Y(0)\}=\pr(A=1|\bmW)=p(\bmW)\in(0,1),
$$
where $p$ is a pre-specified propensity score function. The motivation for considering CDR, a generalization of CIR, is to achieve higher design efficiency \citep{z23}. A CDR design is somewhat similar to an observational study with the important distinction that the propensity score is specified and known by the investigator. In both cases (CIR and CDR), treatment assignment is independent across patients and follows the same mechanism for all patients. The resulting data can be conceptualized as independent copies of $(\bmW,A,Y)$.

Under CIR, efficient estimation of $\delta$ entails making effective use of the information in $\bmW$, for which many methods are available; see Section \ref{intro} for an incomplete list of references. Under CDR, $\delta$ can be estimated using existing methods in the causal inference literature \citep[e.g.,][]{vr03,vr11} coupled with the extra knowledge of the true propensity score; see \citet[Section 3.1]{z23} for a detailed description. For both CIR and CDR, it makes sense to optimize treatment assignment by maximizing the nonparametric information bound. The optimal value of $\pi$ in CIR is
\begin{equation}\label{opt.pi}
\pi_{\text{opt}}=\frac{g'(\mu_1)[\epn\{v_1(\bmW)\}]^{1/2}}{g'(\mu_1)[\epn\{v_1(\bmW)\}]^{1/2}+g'(\mu_0)[\epn\{v_0(\bmW)\}]^{1/2}},
\end{equation}
where $g'$ is the derivative function of $g$ and $v_a(\bmW)=\var\{Y(a)|\bmW\}$, $a=0,1$. The optimal propensity score for CDR is given by
\begin{equation}\label{opt.ps}
p_{\text{opt}}(\bmW)=\frac{g'(\mu_1)v_1(\bmW)^{1/2}}{g'(\mu_1)v_1(\bmW)^{1/2}+g'(\mu_0)v_0(\bmW)^{1/2}}.
\end{equation}
Note that, in both cases, the optimal treatment assignment mechanism depends on the conditional variance functions $(v_1,v_0)$.

To take advantage of these optimality results in designing a trial, one may estimate $(v_1,v_0)$ from preliminary data and substitute the estimates into equation \eqref{opt.pi} or \eqref{opt.ps}. This approach is infeasible if preliminary data are unavailable. Even with preliminary data available, this approach may be unreliable due to high variability and/or potential bias in the preliminary estimates of $(v_1,v_0)$. Indeed, early phase trials tend to be small, and the resulting estimates often lack precision. Sometimes preliminary data come from a confirmatory trial of a similar treatment, in which case one may be more concerned about potential bias (due to treatment differences) than about variability. These difficulties make it challenging to construct a reliable near-optimal design before the trial begins, and motivate our consideration of an adaptive design where we learn from accruing data and refine the treatment assignment mechanism as the trial proceeds.

\section{Adaptive Design}\label{2s.cir}

For ease of presentation, we will focus on a simple two-stage CIR version of the proposed adaptive design in this section and the next one. The design and estimation method are extended to CDR in Appendix A and to multiple stages in Appendix B. The only adaptation we consider here is a possible change in the treatment allocation ratio following the interim analysis; we do not consider other adaptations (e.g., sample size, patient population, or outcome variable) or the possibility of early stopping for success or futility.

Let $(n_1,n_2)$ be pre-specified sample sizes for the two stages. Let $(\bmW_{si},Y_{si}(1),Y_{si}(0))$ denote baseline characteristics for patient $i=1,\dots,n_s$ in stage $s=1,2$; these are assumed to be independent across $(s,i)$ and identically distributed as $(\bmW,Y(1),Y(0))$. Let $A_{si}$ denote the assigned treatment and $Y_{si}=Y_{si}(A_{si})=A_{si}Y_{si}(1)+(1-A_{si})Y_{si}(0)$ the resulting outcome for patient $i$ in stage $s$. The first stage follows a standard CIR design with a pre-specified probability $\pi_1\in(0,1)$ of assigning treatment 1. The resulting data, $\bmD_1=\{\bmO_{1i}=(\bmW_{1i},A_{1i},Y_{1i}),i=1,\dots,n_1\}$, are independent across patients and identically distributed as $\bmO_{1}=(\bmW,A_{1},Y_{1})$, where $\pr\{A_1=1|\bmW,Y(1),Y(0)\}=\pr(A_1=1)=\pi_1$ and $Y_1=Y(A_1)=A_1Y(1)+(1-A_1)Y(0)$.

At the interim analysis, $\bmD_1$ will be analyzed to estimate the marginal means $(\mu_1,\mu_0)$, the conditional variance functions $(v_1,v_0)$, and the optimal allocation $\pi_{\text{opt}}$. Estimating $(\mu_1,\mu_0)$ from $\bmD_1$ is a standard estimation problem for which many methods are available, including augmentation \citep{t08}. To estimate $(v_1,v_0)$ from $\bmD_1$, we note that $v_a(\bmW)=\var(Y_1|A_1=a,\bmW)$. If $\bmW$ is discrete, $v_a(\bmw)$ can be estimated using the sample variance of $\{Y_{1i}:A_{1i}=a,\bmW_{1i}=\bmw\}$ for each $a\in\{0,1\}$ and each possible value $\bmw$ of $\bmW$. If $\bmW$ includes continuous components, estimation of $(v_1,v_0)$ may involve some modeling. For a binary outcome, a logistic regression model for $\pr(Y_1=1|A_1,\bmW)$ would suffice for estimating $(v_1,v_0)$. For other types of outcomes, one could specify models for both $\epn(Y_1|A_1,\bmW)$ and $\var(Y_1|A_1,\bmW)$ and estimate both models using estimating equations \citep[e.g.,][]{dc87}. Once $(\mu_1,\mu_0)$ and $(v_1,v_0)$ are estimated, their estimates can be substituted into equation \eqref{opt.pi} to estimate $\pi_{\text{opt}}$. The resulting estimate of $\pi_{\text{opt}}$, which we denote by $\pi_2$, will be used for treatment assignment in stage 2. If $n_1$ is insufficient for reliable estimation of $(v_1,v_0)$ based on $\bmD_1$, as may be the case if $\bmW$ has many components, the optimization process for choosing $\pi_2$ may be restricted to a coarsened version of $\bmW$, say $\bmX$, which may be a sub-vector or a discrete version of $\bmW$. Operationally, this means replacing $\bmW$ with $\bmX$ in this paragraph as well as equation \eqref{opt.pi}; see \citet[Section 5]{z23} for examples. Whether $\pi_2$ is optimized for all of $\bmW$ or a coarsened version of it has no impact on the subsequent development, where $\pi_2$ is allowed to depend on $\bmD_1$ in an arbitrary, unspecified fashion (subject to regularity conditions).

Once $\pi_2$ is chosen, the trial enters stage 2, which follows a standard CIR design where patients receive treatment 1 with probability $\pi_2$. Given $\pi_2$, the stage 2 data $\bmD_2=\{\bmO_{2i}=(\bmW_{2i},A_{2i},Y_{2i}),i=1,\dots,n_2\}$ are independent and identically distributed as $\bmO_{2}=(\bmW,A_{2},Y_{2})$, where $\pr\{A_2=1|\bmW,Y(1),Y(0)\}=\pr(A_2=1)=\pi_2$ and $Y_2=Y(A_2)=A_2Y(1)+(1-A_2)Y(0)$. Without conditioning on $\pi_2$, the $\bmO_{2i}$'s are identically distributed but not independent of each other. Because $\pi_2$ may depend on $\bmD_1$, the $\bmO_{2i}$'s are generally dependent on $\bmD_1$ as well. Thus, in general, the combined trial data $(\bmD_1,\bmD_2)$ are neither independent nor identically distributed across patients, and standard methods for estimating $\delta$ are not applicable to the current setting.

\section{Treatment Effect Estimation}\label{est.2s.cir}

For each $a\in\{0,1\}$, possible estimators of $\mu_a$ include the stage-specific treatment group averages:
$$
\widehat\mu_a^{(s)}=\frac{\widehat\epn_s\{I(A_{s}=a)Y_{s}\}}{\widehat\epn_s\{I(A_{s}=a)\}},
\qquad s=1,2,
$$
where $\widehat\epn_s$ denotes sample average over $\bmD_s=(\bmO_{s1},\dots,\bmO_{sn_s})$ and $I(\cdot)$ is the indicator function. By conditioning on $(A_{s1},\dots,A_{sn_s})$, it is easy to see that each $\widehat\mu_a^{(s)}$ is unbiased for $\mu_a$. The asymptotic property of $\widehat\mu_a^{(1)}$ is well known, while that of $\widehat\mu_a^{(2)}$ is less transparent due to the randomness of $\pi_2$. The two stage-specific estimators can be combined into a weighted average: $\widehat\mu_a(\theta)=\theta\widehat\mu_a^{(1)}+(1-\theta)\widehat\mu_a^{(2)}$, $\theta\in[0,1]$. Clearly, for any fixed $\theta$, $\widehat\mu_a(\theta)$ remains unbiased for $\mu_a$. The corresponding estimator of $\delta$ is $\widehat\delta(\theta)=g\{\widehat\mu_1(\theta)\}-g\{\widehat\mu_0(\theta)\}$. This estimator is generally inefficient as it does not make use of the available baseline covariate data. To incorporate baseline covariate data for improved efficiency, we consider a class of augmented estimators:
\begin{equation*}
\widehat\delta^{\text{aug}}(b_1,b_2,\theta)=\widehat\delta(\theta)
-\widehat\epn_1\left\{(A_{1}-\pi_1)b_1(\bmW)\right\}-\widehat\epn_2\left\{(A_{2}-\pi_2)b_2(\bmW)\right\},
\end{equation*}
where $b_1$ and $b_2$ are real-valued functions of $\bmW$ such that $\epn\{b_s(\bmW)^2\}<\infty$, $s=1,2$. The following theorem provides the asymptotic distribution of $\widehat\delta^{\text{aug}}(b_1,b_2,\theta)$ with fixed $(b_1,b_2,\theta)$.

\begin{thm}\label{cir.thm1}
Under the two-stage adaptive CIR design described in Section \ref{2s.cir}, assume that, as $n_1\rightarrow\infty$, $n_1/n_2$ converges to some $\lambda\in(0,\infty)$ and $\pi_2$ converges in probability to some $\pi_2^*\in(0,1)$. Then, for any fixed $(b_1,b_2,\theta)$, we have, as $n_1\rightarrow\infty$,
\begin{equation*}\label{cir.asyv}
\sqrt{n_1}\left\{\widehat\delta^{\text{aug}}(b_1,b_2,\theta)-\delta\right\}\stackrel{d}{\longrightarrow} N\left(0, \sigma_{\textsc{cir}}^2(b_1,b_2,\theta)\right),
\end{equation*}
where
\begin{equation*}
\begin{aligned}
\sigma_{\textsc{cir}}^2(b_1,b_2,\theta)&=\var\{\psi_1^{\text{aug}}(\bmO_1)\}+\lambda\var\{\psi_2^{\text{aug}}(\bmO_2^*)\},\\
\psi_1^{\text{aug}}(\bmO_1)&=\frac{\theta g'(\mu_1)A_{1}(Y_{1}-\mu_1)}{\pi_1}-\frac{\theta g'(\mu_0)(1-A_{1})(Y_{1}-\mu_0)}{1-\pi_1}-(A_{1}-\pi_1)b_1(\bmW),\\
\psi_2^{\text{aug}}(\bmO_2^*)&=\frac{(1-\theta)g'(\mu_1)A_{2}^*(Y_{2}^*-\mu_1)}{\pi_2^*}-\frac{(1-\theta)g'(\mu_0)(1-A_{2}^*)(Y_{2}^*-\mu_0)}{1-\pi_2^*}-(A_{2}^*-\pi_2^*)b_2(\bmW),
\end{aligned}
\end{equation*}
and $\bmO_2^*=(\bmW,A_2^*,Y_2^*)$ is similar to $\bmO_2$ with $\pi_2^*$ replacing $\pi_2$:
\begin{equation*}
\begin{gathered}
\pr\{A_2^*=1|\bmW,Y(1),Y(0)\}=\pr(A_2^*=1)=\pi_2^*,\\
Y_2^*=Y(A_2^*)=A_2^*Y(1)+(1-A_2^*)Y(0).
\end{gathered}
\end{equation*}
\end{thm}

The assumption that $\pi_2$ converges to some $\pi_2^*$ is expected to hold if the underlying estimates of $(v_1,v_0)$ (described in Section \ref{2s.cir}) converge to well-defined limits.

It is of interest to minimize $\sigma_{\textsc{cir}}^2(b_1,b_2,\theta)$ with respect to $(b_1,b_2,\theta)$. For any given $\theta$, $\sigma_{\textsc{cir}}^2(b_1,b_2,\theta)$ is minimized by setting $(b_1,b_2)$ equal to
\begin{equation}\label{cir.optb}
\begin{aligned}
b_{1,\text{opt}}(\bmW;\theta)&=\theta\left[\frac{g'(\mu_1)\{m_1(\bmW)-\mu_1\}}{\pi_1}+
\frac{g'(\mu_0)\{m_0(\bmW)-\mu_0\}}{1-\pi_1}\right],\\
b_{2,\text{opt}}(\bmW;\theta)&=(1-\theta)\left[\frac{g'(\mu_1)\{m_1(\bmW)-\mu_1\}}{\pi_2^*}+
\frac{g'(\mu_0)\{m_0(\bmW)-\mu_0\}}{1-\pi_2^*}\right],
\end{aligned}
\end{equation}
where $m_a(\bmW)=\epn\{Y(a)|\bmW\}$, $a=0,1$. A proof of this result is given in Appendix C. The optimal choices $(b_{1,\text{opt}},b_{2,\text{opt}})$ are generally  unknown in reality and need to be estimated from trial data. Suppose $\widetilde{b}_s$ is an estimator of $b_{s,\text{opt}}$ that converges in probability to some limit function $b_s$ in the sense that $\|\widetilde{b}_s-b_s\|_2=o_p(1)$, where $s=1,2$ and $\|h\|_2=[\epn\{h(\bmW)^2\}]^{1/2}$. Under regularity conditions, $\widehat\delta^{\text{aug}}(\widetilde{b}_1,\widetilde{b}_2,\theta)$ is asymptotically normal with asymptotic variance $\sigma_{\textsc{cir}}^2(b_1,b_2,\theta)$ and thus asymptotically equivalent to $\widehat\delta^{\text{aug}}(b_1,b_2,\theta)$ (see Theorem \ref{cir.thm3} below).

Motivated by expression \eqref{cir.optb}, we consider
\begin{equation*}
\begin{aligned}
\widehat b_{1}(\bmW;\theta)&=\theta\left[\frac{g'(\widehat\mu_1)\{\widehat m_1(\bmW)-\widehat\mu_1\}}{\pi_1}+
\frac{g'(\widehat\mu_0)\{\widehat m_0(\bmW)-\widehat\mu_0\}}{1-\pi_1}\right],\\
\widehat b_{2}(\bmW;\theta)&=(1-\theta)\left[\frac{g'(\widehat\mu_1)\{\widehat m_1(\bmW)-\widehat\mu_1\}}{\pi_2}+
\frac{g'(\widehat\mu_0)\{\widehat m_0(\bmW)-\widehat\mu_0\}}{1-\pi_2}\right],
\end{aligned}
\end{equation*}
where $(\widehat\mu_1,\widehat\mu_0)$ and $(\widehat m_1,\widehat m_0)$ are generic estimators of $(\mu_1,\mu_0)$ and $(m_1,m_0)$. Relevant to the estimation of $(m_1,m_0)$ is the fact that $m_a(\bmW)=\epn(Y_1|A_1=a,\bmW)=\epn(Y_2|A_2=a,\bmW)$, $a=0,1$; this holds despite $\bmO_1$ and $\bmO_2$ not being identically distributed. If $\bmW$ is discrete, $m_a(\bmw)$ can be estimated by the sample mean of $\{Y_{i}:A_{i}=a,\bmW_{i}=\bmw\}$ for each $a\in\{0,1\}$ and each possible value $\bmw$ of $\bmW$.  More generally, $(m_1,m_0)$ may be estimated by fitting an outcome regression model to the totality of trial data. We assume that $(\widehat\mu_1,\widehat\mu_0)$ converge to $(\mu_1,\mu_0)$ and $(\widehat m_1,\widehat m_0)$ to some limit functions $(m_1^*,m_0^*)$; see Appendix C for a discussion of this assumption. It follows that $(\widehat b_1,\widehat b_2)$ converge to $(b_1^*,b_2^*)$ defined by
\begin{equation*}
\begin{aligned}
b_{1}^*(\bmW;\theta)&=\theta\left[\frac{g'(\mu_1)\{m_1^*(\bmW)-\mu_1\}}{\pi_1}+
\frac{g'(\mu_0)\{m_0^*(\bmW)-\mu_0\}}{1-\pi_1}\right],\\
b_{2}^*(\bmW;\theta)&=(1-\theta)\left[\frac{g'(\mu_1)\{m_1^*(\bmW)-\mu_1\}}{\pi_2^*}+
\frac{g'(\mu_0)\{m_0^*(\bmW)-\mu_0\}}{1-\pi_2^*}\right].
\end{aligned}
\end{equation*}
For this particular construction, we show in Appendix C that $\sigma_{\textsc{cir}}^2(b_1^*(\cdot;\theta),b_2^*(\cdot;\theta),\theta)$, the asymptotic variance of $\widehat\delta^{\text{aug}}(\widehat b_1(\cdot;\theta),\widehat b_2(\cdot;\theta),\theta)$, is minimized by setting $\theta$ equal to $\theta_{\text{opt}}^*=\lambda\sigma_{2,\textsc{cir}}^{*2}/(\sigma_{1,\textsc{cir}}^{*2}+\lambda\sigma_{2,\textsc{cir}}^{*2})$, where
\begin{equation*}
\begin{aligned}
\sigma_{1,\textsc{cir}}^{*2}&=\var\bigg[\frac{g'(\mu_1)A_1\{Y_1-m_1^*(\bmW)\}}{\pi_1}-\frac{g'(\mu_0)(1-A_1)\{Y_1-m_0^*(\bmW)\}}{1-\pi_1}\\
&\qquad\quad+g'(\mu_1)\left\{m_1^*(\bmW)-\mu_1\right\}-g'(\mu_0)\left\{m_0^*(\bmW)-\mu_0\right\}\bigg],\\
\sigma_{2,\textsc{cir}}^{*2}&=\var\bigg[\frac{g'(\mu_1)A_2^*\{Y_2^*-m_1^*(\bmW)\}}{\pi_2^*}-\frac{g'(\mu_0)(1-A_2^*)\{Y_2^*-m_0^*(\bmW)\}}{1-\pi_2^*}\\
&\qquad\quad+g'(\mu_1)\left\{m_1^*(\bmW)-\mu_1\right\}-g'(\mu_0)\left\{m_0^*(\bmW)-\mu_0\right\}\bigg].\\
\end{aligned}
\end{equation*}

\begin{rmk}\label{cir.eff}
If $(m_1^*,m_0^*)=(m_1,m_0)$, then $\sigma_{1,\textsc{cir}}^{*2}$ is the asymptotic variance of a nonparametric efficient estimator of $\delta$ based on stage 1 data, and $\sigma_{2,\textsc{cir}}^{*2}$ has a similar interpretation pertaining to stage 2 with the random variable $\pi_2$ replaced by its limit $\pi_2^*$. It is easy to see that, with $(m_1^*,m_0^*)=(m_1,m_0)$, the triplet $(b_1^*(\cdot;\theta_{\text{opt}}^*),b_2^*(\cdot;\theta_{\text{opt}}^*),\theta_{\text{opt}}^*)$ is the (essentially) unique minimizer of $\sigma_{\textsc{cir}}^2(b_1,b_2,\theta)$. The global minimum of $\sigma_{\textsc{cir}}^2(b_1,b_2,\theta)$, as a function of the design parameters $(\pi_1,\pi_2^*)$, is minimized when $\pi_1=\pi_2^*=\pi_{\text{opt}}$; this follows from the same argument for proving Theorem 1 of \citet{z23}. Heuristically, the efficiency of the adaptive design in Section \ref{2s.cir} depends on how close $\pi_1$ and $\pi_2^*$ are to the optimal value $\pi_{\text{opt}}$. Compared to a one-stage non-adaptive design that randomizes $n_1+n_2$ patients according to $\pi_1$, the proposed adaptive design is more efficient if $\pi_2^*$ is more efficient than $\pi_1$ in the sense that $\sigma_{2,\textsc{cir}}^{*2}<\sigma_{1,\textsc{cir}}^{*2}$ when $(m_1^*,m_0^*)=(m_1,m_0)$.
\end{rmk}

For any $(m_1^*,m_0^*)$, $\theta_{\text{opt}}^*$ is estimated consistently by $\widehat\theta=n_1\widehat\sigma_{2,\textsc{cir}}^2/(n_2\widehat\sigma_{1,\textsc{cir}}^2+n_1\widehat\sigma_{2,\textsc{cir}}^2)$ with
\begin{equation*}
\begin{aligned}
\widehat\sigma_{s,\textsc{cir}}^2&=\widehat\var_s\bigg[\frac{g'(\widehat\mu_1)A_s\{Y_s-\widehat m_1(\bmW)\}}{\pi_s}-\frac{g'(\widehat\mu_0)(1-A_s)\{Y_s-\widehat m_0(\bmW)\}}{1-\pi_s}\\
&\qquad\qquad+g'(\widehat\mu_1)\left\{\widehat m_1(\bmW)-\widehat\mu_1\right\}-g'(\widehat\mu_0)\left\{\widehat m_0(\bmW)-\widehat\mu_0\right\}\bigg],
\end{aligned}
\end{equation*}
where $\widehat\var_s$ denotes sample variance over $\bmD_s=(\bmO_{s1},\dots,\bmO_{sn_s})$, $s=1,2$. The consistency of $\widehat{\theta}$ is proved in Appendix C. It is natural to substitute $\widehat\theta$ into $\widehat\delta^{\text{aug}}(\widehat b_1(\cdot;\theta),\widehat b_2(\cdot;\theta),\theta)$ and estimate $\delta$ using $\widehat\delta^{\text{aug}}(\widehat b_1(\cdot;\widehat\theta),\widehat b_2(\cdot;\widehat\theta),\widehat\theta)$, whose asymptotic property is given in the next theorem. For brevity, we write $\widehat b_s=\widehat b_s(\cdot;\widehat\theta)$ and $b_s^*=b_s^*(\cdot;\theta_{\text{opt}}^*)$, $s=1,2$.

\begin{thm}\label{cir.thm3}
Assume the conditions in Theorem \ref{cir.thm1} hold. For any random triplet $(\widetilde b_1,\widetilde b_2,\widetilde\theta)$ that element-wise converges in probability to a fixed triplet $(b_1,b_2,\theta)$, we have, as $n_1\rightarrow\infty$,
\begin{equation*}\label{cir.asyv.est}
\sqrt{n_1}\left\{\widehat\delta^{\text{aug}}(\widetilde b_1,\widetilde b_2,\widetilde\theta)-\delta\right\}\stackrel{d}{\longrightarrow} N\left(0, \sigma_{\textsc{cir}}^2(b_1,b_2,\theta)\right);
\end{equation*}
that is, $\widehat\delta^{\text{aug}}(\widetilde b_1,\widetilde b_2,\widetilde\theta)$ is asymptotically equivalent to $\widehat\delta^{\text{aug}}(b_1,b_2,\theta)$. In particular, for $(\widehat b_1,\widehat b_2,\widehat\theta)$ defined in the preceding paragraph, $\widehat\delta^{\text{aug}}(\widehat b_1,\widehat b_2,\widehat\theta)$ is asymptotically equivalent to $\widehat\delta^{\text{aug}}(b_1^*,b_2^*,\theta_{\text{opt}}^*)$ and attains the smallest asymptotic variance among $\{\widehat\delta^{\text{aug}}(\widehat b_1(\cdot;\theta),\widehat b_2(\cdot;\theta),\theta),\theta\in[0,1]\}$. If $(m_1^*,m_0^*)=(m_1,m_0)$, then $\widehat\delta^{\text{aug}}(\widehat b_1,\widehat b_2,\widehat\theta)$ attains the smallest asymptotic variance among all estimators of the form $\widehat\delta^{\text{aug}}(b_1,b_2,\theta)$.
\end{thm}


The asymptotic variance of $\widehat\delta^{\text{aug}}(\widehat b_1,\widehat b_2,\widehat\theta)$, $\sigma_{\textsc{cir}}^2(b_1^*,b_2^*,\theta_{\text{opt}}^*)$, is consistently estimated by
\begin{equation*}
\begin{aligned}
&\widehat\var_1\bigg\{\frac{\widehat{\theta} g'(\widehat\mu_1)A_{1}(Y_{1}-\widehat\mu_1)}{\pi_1}-\frac{\widehat{\theta} g'(\widehat\mu_0)(1-A_{1})(Y_{1}-\widehat\mu_0)}{1-\pi_1}-(A_{1}-\pi_1)\widehat{b}_1(\bmW)\bigg\}\\
&+\frac{n_1}{n_2}\widehat\var_2\bigg\{\frac{(1-\widehat\theta)g'(\widehat\mu_1)A_{2}(Y_{2}-\widehat\mu_1)}{\pi_2}-\frac{(1-\widehat\theta)g'(\widehat\mu_0)(1-A_{2})(Y_{2}-\widehat\mu_0)}{1-\pi_2}-(A_{2}-\pi_2)\widehat{b}_2(\bmW)\bigg\}.
\end{aligned}
\end{equation*}

\section{Simulation}\label{sim}

This section reports a simulation study evaluating the two-stage adaptive CIR and CDR designs together with the proposed estimation methods. In this simulation study, the covariate vector $\bmW=(W_1, W_2, W_3)'$ follows a trivariate normal distribution with mean ${\bf0}$ and a variance matrix whose diagonal elements are all equal to 0.5 and off-diagonal elements equal to 0. We consider a binary outcome which, for each $a\in\{0,1\}$, relates to $\bmW$ through a logistic regression model:
$
\logit[\pr\{Y(a)=1|\bmW\}]=\gamma_0+\gamma_1a+\bmgamma_2'\bmW+\bmgamma_3'(a\bmW),
$
where $\gamma_0=-2.5$, $\gamma_1=1$ or 2, $\bmgamma_2=(-0.2,-0.2,0.2)'$, and $\bmgamma_3=(1,-1,-1.5)'$. The treatment effect of interest is the log-odds ratio
$\delta=\logit[\pr\{Y(1)=1\}]-\logit[\pr\{Y(0)=1\}].$

Under a two-stage design, treatment assignment follows a pre-specified mechanism in stage 1 and an optimized one (based on all available data at the interim analysis) in stage 2. The treatment assignment mechanism for stage 1 may or may not be informed by preliminary data, giving rise to two distinct settings of practical interest:
\begin{description}
\item{Setting 1:} Stage 1 follows the standard 1:1 CIR design with $\pi_1=1/2$;
\item{Setting 2:} Stage 1 follows a CIR or CDR design optimized using preliminary data.
\end{description}
In each setting, for each two-stage design we consider, a one-stage design that utilizes the same stage 1 treatment assignment mechanism throughout the trial is included as a comparator. We set $n_1=n_2=250$ for all two-stage designs and $n=500$ (total sample size) for one-stage designs. Each design is simulated 5,000 times in each distinct situation.

\subsection{Setting 1}\label{sim-sect1}
In Setting 1, we compare three designs: a one-stage 1:1 CIR design, a two-stage CIR design with 1:1 CIR (stage 1) followed by optimized CIR (stage 2), and a two-stage CDR design with 1:1 CIR followed by optimized CDR. In the last two designs, treatment allocation in stage 2 is optimized for $\bmX$, a non-empty sub-vector of $\bmW$, using stage 1 data. For each choice of $\bmX$, the optimal treatment allocation ($\pi_{\text{opt}}$ or $p_{\text{opt}}$) is estimated by substituting estimates of $(\mu_1,\mu_0)$ and $(v_1,v_0)$ from stage 1 data into equation \eqref{opt.pi} or \eqref{opt.ps}. Stage 1 estimates of $(v_1,v_0)$ are obtained by fitting the following logistic regression model:
\begin{equation}\label{sim.model.condPara}
\logit\{\pr(Y_1=1|A_1,\bmX)\}=\alpha_0+\alpha_1A_1+\bmalpha_2'\bmX+\bmalpha_3'(A_1\bmX).
\end{equation}
This working model is misspecified unless $\bmX=\bmW$.

Under each design, $\delta$ is estimated using two estimators: a simple one and an optimized one. For the two-stage CIR design, the simple estimator is $\widehat\delta(\theta_0)$, where $\theta_0=n_1/(n_1+n_2)=1/2$, and the optimized estimator is $\widehat\delta^{\text{aug}}(\widehat b_1,\widehat b_2,\widehat\theta)$. As described in Section \ref{est.2s.cir}, $(\widehat b_1,\widehat b_2)$ are based on $(\widehat m_1,\widehat m_0)$, which in this case are obtained by fitting a logistic regression model similar to \eqref{sim.model.condPara}, with $\bmX$ replaced by $\bmW$, to all trial data (from both stages). For the two-stage CDR design, the simple estimator is $\widehat\delta^{\text{ipw}}(\eta_0)$, where $\eta_0=\theta_0$, and the optimized estimator is $\widehat\delta^{\text{aipw}}(\widehat c_1,\widehat c_2,\widehat\eta)$, where $(\widehat c_1,\widehat c_2)$ are based on the same $(\widehat m_1,\widehat m_0)$ mentioned earlier (both estimators are described in Appendix A). For the one-stage design, the simple estimator is based on treatment-specific averages and the optimized one is an augmented estimator described in \citet[Section 2]{z23}. 

As expected, estimation bias is negligible for all estimators and all designs (results not shown). Table \ref{sim.rst.re} reports relative efficiency results with the simple estimator under the one-stage design serving as the reference. The relative efficiency of an estimator is calculated as the inverse ratio of its empirical variance to that of the reference estimator. Under each design, the optimized estimator is clearly advantageous to the simple estimator. Comparing the optimized estimators for the three designs, the two-stage CIR design is consistently more efficient than the one-stage design, and the two-stage CDR design is even more efficient. Apparently, the stage 1 data with $n_1=250$ provide adequate support for estimating the optimal CDR design. For the optimized estimators under the two-stage designs, Table S1 in Appendix C indicates that the proposed variance estimators are virtually median-unbiased, and Table \ref{sim.rst.cp} demonstrates that the resulting confidence intervals provide close-to-nominal coverage.

\subsection{Setting 2}\label{sim-sect2}

In Setting 2, we compare five different designs: a one-stage CIR design with optimized CIR, a one-stage CDR design with optimized CDR, a two-stage CIR design with optimized CIR (stage 1) followed by further optimized CIR (stage 2), a two-stage CDR design with optimized CDR followed by further optimized CDR, and a two-stage hybrid design with optimized CIR followed by optimized CDR. All attempts to optimize treatment allocation make use of a preliminary dataset of size $n_0=100$, generated under 1:1 CIR in the same patient population described earlier. (The optimizations for stage 2 also make use of stage 1 data.) As in Setting 1, each optimization is based on a sub-vector $\bmX$ of $\bmW$ and involves fitting the logistic regression model \eqref{sim.model.condPara} to the available data.

Under each design, $\delta$ is estimated using a simple estimator and an optimized one.  For the one-stage CIR design and the two-stage designs, these are the same estimators described and compared in Section \ref{sim-sect1}. (The two-stage hybrid design is approached as a two-stage CDR design for the purpose of treatment effect estimation.) For the one-stage CDR design, the simple estimator is an IPW estimator and the optimized estimator is an AIPW estimator; both are described in \citet[Section 3.1]{z23}.

The relative efficiency results, with the optimized estimator under the one-stage CIR design as the reference, are shown in the lower section of Table \ref{sim.rst.re}.  The one-stage CDR design generally underperforms the one-stage CIR design, suggesting that the preliminary data with $n_0=100$ may be insufficient to support full optimization of CDR. The two-stage CIR and CDR designs perform similarly to each other, and both designs generally outperform the one-stage designs. The two-stage hybrid design appears to perform even better than the two-stage CIR and CDR designs, though the incremental improvement tends to be small. To understand the superior performance of the two-stage hybrid design, we note that its first stage (optimized CIR) acknowledges the limited amount ($n_0=100$) of preliminary data and its second stage (optimized CDR) takes full advantage of the combined data ($n_0+n_1=350$) available at the interim. Tables \ref{sim.rst.cp} and S1 demonstrate adequate coverage and variance estimation for the two-stage optimized estimators.


\subsection{Additional Simulation Results}

Appendix D reports the results of additional simulation studies with various deviations from the main study described here: a smaller sample size ($n=300$), different sample size allocations between stages, inclusion of non-prognostic covariates, and more severe misspecification of the working model for interim estimation of $(v_1,v_0)$. The results are generally consistent with those reported here, with a few new insights (detailed in Appendix D).

\section{Application}\label{app}

As an illustration, this section describes how the proposed methodology can be used to re-design and re-analyze the NINDS rt-PA trial introduced in Section \ref{intro}. Following previous analyses of the same trial \citep{n95,i04,zm19}, we will compare treatments at 3 months post-randomization with respect to four clinical outcomes: Barthel Index, Modified Rankin Scale, Glasgow Outcome Scale, and National Institutes of Health Stroke Scale (NIHSS). For ease of interpretation, each outcome is linearly re-scaled to the unit interval in such a way that higher values are desirable. For each outcome, the efficacy of rt-PA relative to placebo is measured by the between-group difference (rt-PA minus placebo) in the re-scaled mean outcome (i.e., $\delta=\mu_1-\mu_0$).

Our objective is to improve treatment assignment for statistical efficiency by incorporating baseline covariates. For this purpose, we consider three baseline covariates: age, baseline NIHSS, and history of diabetes. History of diabetes is inherently dichotomous, and the other two covariates are dichotomized at their respective medians. The three dichotomous covariates together comprise the covariate vector $\bmW$. 
As alternatives to the original 1:1 CIR design, we consider a two-stage CIR design (1:1 CIR followed by optimized CIR) and a two-stage CDR design (1:1 CIR followed by optimized CDR). Without actual enrollment information, we randomly partition the trial participants into two stages in a 1:1 ratio. We use stage 1 data to optimize treatment allocation in stage 2 based on a covariate vector $\bmX$, which consists of one or more of the three dichotomous covariates described earlier. The conditional variance functions $(v_1,v_0)$ are estimated empirically, as indicated in Section \ref{2s.cir}. For each patient in stage 2, $A_{2i}$ is assigned randomly according to the optimized CIR or CDR design, and $Y_{2i}$ is chosen randomly from the observed outcome values among the actual trial participants in the same treatment group with the same covariate value. Only one set of trial data is generated for each design.
We estimate $\delta$ using optimized estimators that are virtually identical to those described in Section \ref{sim-sect1} with two minor differences. First, the estimand here is a mean difference rather than a log-odds ratio. Second, the discrete nature of the current $\bmW$ allows $(m_1,m_0)$ to be estimated empirically using sample averages. 

Table \ref{ex.rst}  reports the results of this analysis (point estimates of $\delta$, standard errors, and relative efficiency results) for the four clinical outcomes and different choices of $\bmX$. The relative efficiency is calculated as a ratio of variance estimates, with the one-stage CIR design as the reference. As expected, standard errors are generally smaller for the two-stage designs as compared to the one-stage design. The reductions in standard errors may be small on the absolute scale but can be substantial on the relative efficiency scale. Point estimates are typically larger under the two-stage designs as compared to the one-stage design. Taken together, the results in Table \ref{ex.rst} clearly indicate that the use of rt-PA has a beneficial effect, although its statistical significance is difficult to quantify due to multiplicity issues.

\section{Discussion}\label{disc}


An important practical question in applying the proposed design is how to choose stage-specific sample sizes. For example, in a two-stage design with a fixed total sample size, increasing $n_1$ improves estimation precision in the interim analysis but reduces the impact of the estimated optimal design (due to decreased $n_2$). We have investigated this trade-off in a simulation study reported in Appendix D, and the results suggest that the optimal allocation between stages might be some intermediate value. Further research is needed to better characterize that optimal allocation and provide helpful guidance to practitioners.

In this article, we have focused on treatment assignment as the only adaptable aspect of the trial, without considering other possible adaptations (e.g., sample size re-estimation). It is of interest to consider how to make the design more flexible by incorporating other possible adaptations while maintaining the integrity of the trial. 


\section*{Acknowledgements}

The research of Wei Zhang was supported by the National Key R\&D Program of China (2022YFA1004800). The research of Aiyi Liu was supported by the intramural research program of the \textit{Eunice Kennedy Shriver} National Institute of Child Health and Human Development.

\section*{Supporting Information}
Web Appendices and Tables referenced in Sections 3-5, and data and code referenced in Section 5 are available with this paper at the Biometrics website on Oxford Academic.

\section*{Data Availability Statement}

The NINDS rt-PA trial data are available from the corresponding
author upon reasonable request.

\pagebreak
\renewcommand{\baselinestretch}{1.1}
\begin{table}[H]
{\footnotesize
\caption{Simulation-based relative efficiency results for simple and optimized estimators under various one-stage (1S) and two-stage (2S) designs (see Section \ref{sim} for details).}\label{sim.rst.re}
\newcolumntype{d}{D{.}{.}{2}}
\newcolumntype{e}{D{.}{.}{1}}
\begin{center}
\begin{tabular}{ccccddddddd}
\hline
\hline
Setting&$\gamma_1$&Design&Estimator&\multicolumn{7}{c}{$\bmX$}\\
\cline{5-11}
&&&&\multicolumn{1}{c}{$W_1$}&\multicolumn{1}{c}{$W_2$}&\multicolumn{1}{c}{$W_3$}&\multicolumn{1}{c}{$(W_1,W_2)'$}&\multicolumn{1}{c}{$(W_1,W_3)'$}&\multicolumn{1}{c}{$(W_2,W_3)'$}&\multicolumn{1}{c}{$\bmW$}\\
\hline
 1&1&1S CIR&simple&0.95&0.95&0.95&0.95&0.95&0.95&0.95\\
      &&&optimized&1.00&1.00&1.00&1.00&1.00&1.00&1.00\\
\cline{3-11}
   &&2S CIR&simple&0.98&0.99&0.98&0.99&0.98&0.98&0.99\\
      &&&optimized&1.11&1.12&1.12&1.12&1.12&1.13&1.14\\
\cline{3-11}
   &&2S CDR&simple&1.05&1.03&1.05&1.03&1.03&0.98&0.91\\
      &&&optimized&1.16&1.15&1.18&1.16&1.18&1.18&1.18\\
\cline{2-11}
  &2&1S CIR&simple&0.95&0.95&0.95&0.95&0.95&0.95&0.95\\
      &&&optimized&1.00&1.00&1.00&1.00&1.00&1.00&1.00\\
\cline{3-11}
   &&2S CIR&simple&0.99&1.00&0.99&1.00&0.99&1.00&1.00\\
      &&&optimized&1.12&1.13&1.13&1.13&1.13&1.14&1.15\\
\cline{3-11}
   &&2S CDR&simple&1.05&1.02&1.03&1.03&1.01&0.97&0.88\\
      &&&optimized&1.16&1.15&1.16&1.17&1.16&1.17&1.18\\
\hline
 2&1&1S CIR&simple&0.87&0.87&0.88&0.87&0.86&0.86&0.84\\
      &&&optimized&1.00&1.00&1.00&1.00&1.00&1.00&1.00\\
\cline{3-11}
   &&1S CDR&simple&0.81&0.82&0.81&0.77&0.75&0.65&0.42\\
      &&&optimized&0.88&0.77&0.67&0.79&0.91&0.67&0.31\\
\cline{3-11}
   &&2S CIR&simple&0.90&0.91&0.91&0.86&0.85&0.87&0.88\\
      &&&optimized&1.04&1.06&1.05&1.01&1.01&1.04&1.05\\
\cline{3-11}
   &&2S CDR&simple&0.87&0.86&0.90&0.83&0.82&0.71&0.57\\
      &&&optimized&1.00&1.01&1.04&1.00&1.01&0.99&0.97\\
\cline{3-11}
&&2S Hybrid&simple&0.92&0.90&0.93&0.86&0.90&0.82&0.77\\
      &&&optimized&1.05&1.06&1.09&1.01&1.06&1.05&1.09\\
\cline{2-11}
  &2&1S CIR&simple&0.91&0.87&0.88&0.88&0.87&0.87&0.85\\
      &&&optimized&1.00&1.00&1.00&1.00&1.00&1.00&1.00\\
\cline{3-11}
   &&1S CDR&simple&0.81&0.81&0.78&0.74&0.74&0.66&0.45\\
      &&&optimized&0.90&0.89&0.70&0.81&0.48&0.54&0.70\\
\cline{3-11}
   &&2S CIR&simple&0.92&0.87&0.87&0.88&0.89&0.87&0.87\\
      &&&optimized&1.01&1.01&1.02&1.03&1.06&1.05&1.03\\
\cline{3-11}
   &&2S CDR&simple&0.84&0.83&0.84&0.81&0.82&0.75&0.57\\
      &&&optimized&0.99&0.99&1.01&1.00&1.03&1.04&0.96\\
\cline{3-11}
&&2S Hybrid&simple&0.89&0.86&0.87&0.89&0.92&0.85&0.73\\
      &&&optimized&1.02&1.02&1.02&1.05&1.09&1.09&1.06\\
\hline
\end{tabular}
\end{center}
}
\end{table}

\pagebreak
\renewcommand{\baselinestretch}{1.1}
\begin{table}[H]
{\small
\caption{Simulation-based coverage results (at nominal level 95\%) for optimized estimators under various two-stage (2S) designs (see Section \ref{sim} for details).}\label{sim.rst.cp}
\newcolumntype{d}{D{.}{.}{3}}
\newcolumntype{e}{D{.}{.}{1}}
\begin{center}
\begin{tabular}{cccddddddd}
\hline
\hline
Setting&$\gamma_1$&Design&\multicolumn{7}{c}{$\bmX$}\\
\cline{4-10}
&&&\multicolumn{1}{c}{$W_1$}&\multicolumn{1}{c}{$W_2$}&\multicolumn{1}{c}{$W_3$}&\multicolumn{1}{c}{$(W_1,W_2)'$}&\multicolumn{1}{c}{$(W_1,W_3)'$}&\multicolumn{1}{c}{$(W_2,W_3)'$}&\multicolumn{1}{c}{$\bmW$}\\
\hline
1&1&2S CIR  &0.949& 0.950&0.951&0.952&0.952&0.951&0.954\\
  &&2S CDR  &0.957& 0.956&0.954&0.956&0.959&0.957&0.958\\
\cline{2-10}
&2&2S CIR  &0.951& 0.949&0.951&0.950&0.951&0.952&0.951\\
 &&2S CDR  &0.953& 0.952&0.951&0.956&0.954&0.955&0.957\\
\hline
2&1&2S CIR  &0.953& 0.954&0.951&0.947&0.949&0.956&0.951\\
  &&2S CDR  &0.944& 0.951&0.950&0.954&0.951&0.952&0.950\\
  &&2S Hybrid &0.955& 0.953&0.953&0.950&0.952&0.953&0.958\\
\cline{2-10}
&2&2S CIR  &0.948& 0.953&0.952&0.949&0.949&0.952&0.951\\
 &&2S CDR  &0.948& 0.953&0.955&0.954&0.956&0.953&0.946\\
 &&2S Hybrid  &0.953& 0.953&0.955&0.954&0.952&0.957&0.954\\
\hline
\end{tabular}
\end{center}
}
\end{table}

\pagebreak
\renewcommand{\baselinestretch}{1.1}
\begin{table}[H]
{\scriptsize
\caption{Results of re-designing and re-analyzing the NINDS trial: point estimates (standard errors; relative efficiency) of the re-scaled mean score difference for various choices of $Y$ and $\bmX$, under the original one-stage (1S) 1:1 CIR and proposed two-stage (2S) CIR and CDR designs.}\label{ex.rst}
\newcolumntype{d}{D{.}{.}{3}}
\newcolumntype{e}{D{.}{.}{1}}
\begin{center}
\begin{tabular}{cccccc}
\hline
\hline
$\bmX$&\multicolumn{1}{c}{1S CIR}&&\multicolumn{1}{c}{2S CIR}&&\multicolumn{1}{c}{2S CDR}\\
\hline
\multicolumn{6}{c}{$Y=\text{Barthel Index}$}\\
age                  &0.096 (0.029; ref)&&0.125 (0.029; 1.03)&&0.136 (0.028; 1.06)\\
NIHSS                &0.096 (0.029; ref)&&0.134 (0.029; 1.04)&&0.092 (0.028; 1.14)\\
diabetes             &0.096 (0.029; ref)&&0.115 (0.029; 1.04)&&0.109 (0.028; 1.11)\\
age, NIHSS           &0.096 (0.029; ref)&&0.079 (0.029; 1.03)&&0.110 (0.029; 1.02)\\
age, diabetes        &0.096 (0.029; ref)&&0.107 (0.029; 1.06)&&0.119 (0.029; 1.00)\\
NIHSS, diabetes      &0.096 (0.029; ref)&&0.103 (0.029; 1.05)&&0.126 (0.029; 1.04)\\
age, NIHSS, diabetes &0.096 (0.029; ref)&&0.088 (0.028; 1.07)&&0.138 (0.029; 1.04)\\
\hline
\multicolumn{6}{c}{$Y=\text{Modified Rankin Scale}$}\\
age                  &0.095 (0.024; ref)&&0.106 (0.024; 1.02)&&0.117 (0.024; 1.00)\\
NIHSS                &0.095 (0.024; ref)&&0.115 (0.024; 1.01)&&0.116 (0.023; 1.13)\\
diabetes             &0.095 (0.024; ref)&&0.132 (0.024; 1.03)&&0.118 (0.024; 1.01)\\
age, NIHSS           &0.095 (0.024; ref)&&0.084 (0.024; 1.03)&&0.095 (0.024; 1.06)\\
age, diabetes        &0.095 (0.024; ref)&&0.107 (0.024; 1.01)&&0.136 (0.024; 1.04)\\
NIHSS, diabetes      &0.095 (0.024; ref)&&0.117 (0.024; 1.05)&&0.124 (0.024; 1.03)\\
age, NIHSS, diabetes &0.095 (0.024; ref)&&0.096 (0.023; 1.07)&&0.098 (0.024; 1.00)\\
\hline
\multicolumn{6}{c}{$Y=\text{Glasgow Outcome Scale}$}\\
age                  &0.066 (0.021; ref)&&0.089 (0.021; 1.01)&&0.082 (0.021; 1.05)\\
NIHSS                &0.066 (0.021; ref)&&0.070 (0.021; 1.01)&&0.087 (0.021; 1.04)\\
diabetes             &0.066 (0.021; ref)&&0.088 (0.021; 1.02)&&0.087 (0.021; 1.04)\\
age, NIHSS           &0.066 (0.021; ref)&&0.078 (0.021; 1.01)&&0.072 (0.020; 1.11)\\
age, diabetes        &0.066 (0.021; ref)&&0.092 (0.020; 1.09)&&0.090 (0.021; 1.03)\\
NIHSS, diabetes      &0.066 (0.021; ref)&&0.075 (0.020; 1.11)&&0.111 (0.021; 1.02)\\
age, NIHSS, diabetes &0.066 (0.021; ref)&&0.079 (0.020; 1.05)&&0.089 (0.021; 1.01)\\
\hline
\multicolumn{6}{c}{$Y=\text{NIHSS}$}\\
age                  &0.060 (0.027; ref)&&0.067 (0.026; 1.04)&&0.066 (0.027; 0.98)\\
NIHSS                &0.060 (0.027; ref)&&0.059 (0.024; 1.18)&&0.106 (0.025; 1.13)\\
diabetes             &0.060 (0.027; ref)&&0.035 (0.025; 1.09)&&0.076 (0.026; 1.08)\\
age, NIHSS           &0.060 (0.027; ref)&&0.079 (0.026; 1.06)&&0.114 (0.026; 1.05)\\
age, diabetes        &0.060 (0.027; ref)&&0.081 (0.027; 0.98)&&0.097 (0.026; 1.07)\\
NIHSS, diabetes      &0.060 (0.027; ref)&&0.079 (0.026; 1.01)&&0.064 (0.026; 1.07)\\
age, NIHSS, diabetes &0.060 (0.027; ref)&&0.044 (0.026; 1.06)&&0.065 (0.026; 1.03)\\
\hline
\end{tabular}
\end{center}
}
\end{table}

\pagebreak
\renewcommand{\thetable}{S\arabic{table}}
\setcounter{table}{0}
\renewcommand{\thethm}{S\arabic{thm}}
\setcounter{thm}{0}

\begin{center}
{\LARGE Supplementary Materials for \lq\lq An Adaptive Design for Optimizing Treatment Assignment in Randomized Clinical Trials" by Wei Zhang$^{1}$, Zhiwei Zhang$^{2,*}$, and Aiyi Liu$^{3}$}
\end{center}
\begin{center}
$^1$State Key Laboratory of Mathematical Sciences, Academy of Mathematics and Systems Science, Chinese Academy of Sciences, Beijing, China\\
$^2$Biostatistics Innovation Group, Gilead Sciences, Foster City, California, USA\\
$^3$Biostatistics and Bioinformatics Branch, Division of Population Health Research, \textit{Eunice Kennedy Shriver} National Institute of Child Health and Human Development, National Institutes of Health, Bethesda, Maryland, USA\\
$^*$Zhiwei.Zhang6@gilead.com
\end{center}

\vspace{0.5cm}

\section*{Appendix A: Extension to  CDR}\label{2s.cdr}

In this appendix, we extend the adaptive design in Section 3 and the estimation method in Section 4 to the CDR setting. We take Sections 3 and 4 as background information and focus attention on important differences due to CDR. Unless otherwise stated, the notations in Sections 3 and 4 remain applicable in the present setting.

Let $p_1(\bmW)$ be the pre-specified propensity score for treatment assignment in stage 1. The stage 1 data, $\bmD_1=\{\bmO_{1i}=(\bmW_{1i},A_{1i},Y_{1i}),i=1,\dots,n_1\}$, are independent across patients and identically distributed as $\bmO_{1}=(\bmW,A_{1},Y_{1})$, where $\pr\{A_1=1|\bmW,Y(1),Y(0)\}=\pr(A_1=1|\bmW)=p_1(\bmW)$ and $Y_1=Y(A_1)=A_1Y(1)+(1-A_1)Y(0)$. At the interim analysis, $\bmD_1$ will be used to estimate $(\mu_1,\mu_0)$, $(v_1,v_0)$, and the optimal propensity score $p_{\text{opt}}$. Estimation of $(\mu_1,\mu_0)$ in a single-stage CDR trial is discussed in \citet[Section 3.1]{z23}. Estimation of $(v_1,v_0)$ follows the same considerations given in Section 3 because CDR implies $v_a(\bmW)=\var(Y_1|A_1=a,\bmW)$, $a=0,1$. Estimates of $(\mu_1,\mu_0)$ and $(v_1,v_0)$ can be substituted into equation (2) to obtain an estimate of $p_{\text{opt}}$, say $p_2$, which will then be used as the propensity score function in stage 2. Given $p_2$, the stage 2 data $\bmD_2=\{\bmO_{2i}=(\bmW_{2i},A_{2i},Y_{2i}),i=1,\dots,n_2\}$ are independent and identically distributed as $\bmO_{2}=(\bmW,A_{2},Y_{2})$, where $\pr\{A_2=1|\bmW,Y(1),Y(0)\}=\pr(A_2=1|\bmW)=p_2(\bmW)$ and $Y_2=Y(A_2)=A_2Y(1)+(1-A_2)Y(0)$. As noted in Section 3, the combined trial data $(\bmD_1,\bmD_2)$ are neither independent nor identically distributed across patients.

The rest of this section is focused on treatment effect estimation. To account for CDR, we replace the initial stage-specific estimators $\widehat\mu_a^{(s)}$ with inverse probability weighted (IPW) estimators:
$$
\widehat\mu_a^{\text{ipw}(s)}=\frac{\widehat\epn_s\Big(I(A_{s}=a)Y_{s}\big/\big[p_s(\bmW)^a\{1-p_s(\bmW)\}^{1-a}\big]\Big)}{\widehat\epn_s\Big(I(A_{s}=a)\big/\big[p_s(\bmW)^a\{1-p_s(\bmW)\}^{1-a}\big]\Big)},
\qquad s=1,2,\qquad a=0,1.
$$
These can be combined as in Section 4 to yield $\widehat\mu_a^{\text{ipw}}(\eta)=\eta\widehat\mu_a^{\text{ipw}(1)}+(1-\eta)\widehat\mu_a^{\text{ipw}(2)}$, $a=0,1$, and $\widehat\delta^{\text{ipw}}(\eta)=g\{\widehat\mu_1^{\text{ipw}}(\eta)\}-g\{\widehat\mu_0^{\text{ipw}}(\eta)\}$, $\eta\in[0,1]$. For improved efficiency, we consider a class of augmented IPW (AIPW) estimators of $\delta$:
\begin{equation*}
\widehat\delta^{\text{aipw}}(c_1,c_2,\eta)=\widehat\delta^{\text{ipw}}(\eta)
-\widehat\epn_1\left[\{A_{1}-p_1(\bmW)\}c_1(\bmW)\right]-\widehat\epn_2\left[\{A_{2}-p_2(\bmW)\}c_2(\bmW)\right],
\end{equation*}
where $c_1$ and $c_2$ are real-valued functions of $\bmW$ such that $\epn\{c_s(\bmW)^2\}<\infty$, $s=1,2$. The asymptotic distribution of $\widehat\delta^{\text{aipw}}(c_1,c_2,\eta)$ with fixed $(c_1,c_2,\eta)$ is given in the next theorem, which is analogous to Theorem 1.

\begin{thm}\label{cdr.thm5}
Under the two-stage adaptive CDR design described in this section, assume that, as $n_1\rightarrow\infty$, $n_1/n_2$ converges to some $\lambda\in(0,\infty)$ and $\|p_2-p_2^*\|_2=o_p(1)$ for some real-valued function $p_2^*$ such that $\epn\{p_2^*(\bmW)^2\}<\infty$. Then, for any fixed $(c_1,c_2,\eta)$, we have, as $n_1\rightarrow\infty$,
\begin{equation}\label{cdr.asyv}
\sqrt{n_1}\left\{\widehat\delta^{\text{aipw}}(c_1,c_2,\eta)-\delta\right\}\stackrel{d}{\longrightarrow} N\left(0, \sigma_{\textsc{cdr}}^2(c_1,c_2,\eta)\right),
\end{equation}
where
\begin{equation*}
\begin{aligned}
\sigma_{\textsc{cdr}}^2(c_1,c_2,\eta)&=\var\left\{\psi_1^{\text{aipw}}(\bmO_1)\right\}+\lambda\var\left\{\psi_2^{\text{aipw}}(\bmO_2^*)\right\},\\
\psi_1^{\text{aipw}}(\bmO_1)&=\frac{\eta g'(\mu_1)A_{1}(Y_{1}-\mu_1)}{p_1(\bmW)}-\frac{\eta g'(\mu_0)(1-A_{1})(Y_{1}-\mu_0)}{1-p_1(\bmW)}-\{A_{1}-p_1(\bmW)\}c_1(\bmW),\\
\psi_2^{\text{aipw}}(\bmO_2^*)&=\frac{(1-\eta)g'(\mu_1)A_{2}^*(Y_{2}^*-\mu_1)}{p_2^*(\bmW)}-\frac{(1-\eta)g'(\mu_0)(1-A_{2}^*)(Y_{2}^*-\mu_0)}{1-p_2^*(\bmW)}\\
&\quad-\{A_{2}^*-p_2^*(\bmW)\}c_2(\bmW),
\end{aligned}
\end{equation*}
and $\bmO_2^*=(\bmW,A_2^*,Y_2^*)$ is similar to $\bmO_2$ with $p_2^*$ replacing $p_2$:
\begin{equation*}
\begin{gathered}
\pr\{A_2^*=1|\bmW,Y(1),Y(0)\}=\pr(A_2^*=1|\bmW)=p_2^*(\bmW),\\
Y_2^*=Y(A_2^*)=A_2^*Y(1)+(1-A_2^*)Y(0).
\end{gathered}
\end{equation*}
\end{thm}

The asymptotic variance $\sigma_{\textsc{cdr}}^2(c_1,c_2,\eta)$ can be minimized in two steps, in a similar fashion to the minimization of $\sigma_{\textsc{cir}}^2(b_1,b_2,\theta)$ in Section 4. We show in Appendix C that, for any given $\eta$, $\sigma_{\textsc{cdr}}^2(c_1,c_2,\eta)$ is minimized by setting $(c_1,c_2)$ equal to
\begin{equation*}\label{cdr.optc}
\begin{aligned}
c_{1,\text{opt}}(\bmW;\eta)&=\eta\left[\frac{g'(\mu_1)\{m_1(\bmW)-\mu_1\}}{p_1(\bmW)}+
\frac{g'(\mu_0)\{m_0(\bmW)-\mu_0\}}{1-p_1(\bmW)}\right],\\
c_{2,\text{opt}}(\bmW;\eta)&=(1-\eta)\left[\frac{g'(\mu_1)\{m_1(\bmW)-\mu_1\}}{p_2^*(\bmW)}+
\frac{g'(\mu_0)\{m_0(\bmW)-\mu_0\}}{1-p_2^*(\bmW)}\right],
\end{aligned}
\end{equation*}
where $m_a(\bmW)=\epn\{Y(a)|\bmW\}=\epn(Y_s|A_s=a,\bmW)$, $a=0,1$, $s=1,2$. These optimal choices are estimated by
\begin{equation*}
\begin{aligned}
\widehat c_{1}(\bmW;\eta)&=\eta\left[\frac{g'(\widehat\mu_1)\{\widehat m_1(\bmW)-\widehat \mu_1\}}{p_1(\bmW)}+
\frac{g'(\widehat\mu_0)\{\widehat m_0(\bmW)-\widehat\mu_0\}}{1-p_1(\bmW)}\right],\\
\widehat c_{2}(\bmW;\eta)&=(1-\eta)\left[\frac{g'(\widehat\mu_1)\{\widehat m_1(\bmW)-\widehat\mu_1\}}{p_2(\bmW)}+
\frac{g'(\widehat\mu_0)\{\widehat m_0(\bmW)-\widehat\mu_0\}}{1-p_2(\bmW)}\right],
\end{aligned}
\end{equation*}
where $(\widehat\mu_1,\widehat\mu_0)$ and $(\widehat m_1,\widehat m_0)$ are generic estimators of $(\mu_1,\mu_0)$ and $(m_1,m_0)$. We assume that $(\widehat\mu_1,\widehat\mu_0)$ converge to $(\mu_1,\mu_0)$ and $(\widehat m_1,\widehat m_0)$ to some limit functions $(m_1^*,m_0^*)$; then $(\widehat c_1,\widehat c_2)$ converge to $(c_1^*,c_2^*)$ defined by
\begin{equation*}
\begin{aligned}
c_{1}^*(\bmW;\eta)&=\eta\left[\frac{g'(\mu_1)\{m_1^*(\bmW)-\mu_1\}}{p_1(\bmW)}+
\frac{g'(\mu_0)\{m_0^*(\bmW)-\mu_0\}}{1-p_1(\bmW)}\right],\\
c_{2}^*(\bmW;\eta)&=(1-\eta)\left[\frac{g'(\mu_1)\{m_1^*(\bmW)-\mu_1\}}{p_2^*(\bmW)}+
\frac{g'(\mu_0)\{m_0^*(\bmW)-\mu_0\}}{1-p_2^*(\bmW)}\right].
\end{aligned}
\end{equation*}
We show in Appendix C that $\sigma_{\textsc{cdr}}^2(c_1^*(\cdot;\eta),c_2^*(\cdot;\eta),\eta)$ is minimized by setting $\eta$ equal to $\eta_{\text{opt}}^*=\lambda\sigma_{2,\textsc{cdr}}^{*2}/(\sigma_{1,\textsc{cdr}}^{*2}+\lambda\sigma_{2,\textsc{cdr}}^{*2})$, where
\begin{equation*}
\begin{aligned}
\sigma_{1,\textsc{cdr}}^{*2}&=\var\bigg[\frac{g'(\mu_1)A_1\{Y_1-m_1^*(\bmW)\}}{p_1(\bmW)}-\frac{g'(\mu_0)(1-A_1)\{Y_1-m_0^*(\bmW)\}}{1-p_1(\bmW)}\\
&\qquad\quad+g'(\mu_1)\left\{m_1^*(\bmW)-\mu_1\right\}-g'(\mu_0)\left\{m_0^*(\bmW)-\mu_0\right\}\bigg],\\
\sigma_{2,\textsc{cdr}}^{*2}&=\var\bigg[\frac{g'(\mu_1)A_2^*\{Y_2^*-m_1^*(\bmW)\}}{p_2^*(\bmW)}-\frac{g'(\mu_0)(1-A_2^*)\{Y_2^*-m_0^*(\bmW)\}}{1-p_2^*(\bmW)}\\
&\qquad\quad+g'(\mu_1)\left\{m_1^*(\bmW)-\mu_1\right\}-g'(\mu_0)\left\{m_0^*(\bmW)-\mu_0\right\}\bigg].\\
\end{aligned}
\end{equation*}
Remark 1 is easily extensible to the present setting; for example, $\sigma_{\textsc{cdr}}^2(c_1,c_2,\eta)$ is minimized by the triplet $(c_1^*(\cdot;\eta_{\text{opt}}^*),c_2^*(\cdot;\eta_{\text{opt}}^*),\eta_{\text{opt}}^*)$ with $(m_1^*,m_0^*)=(m_1,m_0)$, and the minimum value of $\sigma_{\textsc{cdr}}^2(c_1,c_2,\eta)$, as a function of the design parameters $(p_1,p_2^*)$, is minimized when $p_1=p_2^*=p_{\text{opt}}$. Regardless of $(m_1^*,m_0^*)$, $\eta_{\text{opt}}^*$ is estimated consistently by $\widehat\eta=n_1\widehat\sigma_{2,\textsc{cdr}}^2/(n_2\widehat\sigma_{1,\textsc{cdr}}^2+n_1\widehat\sigma_{2,\textsc{cdr}}^2)$ with
\begin{equation*}
\begin{aligned}
\widehat\sigma_{s,\textsc{cdr}}^2&=\widehat\var_s\bigg[\frac{g'(\widehat\mu_1)A_s\{Y_s-\widehat m_1(\bmW)\}}{p_s(\bmW)}-\frac{g'(\widehat\mu_0)(1-A_s)\{Y_s-\widehat m_0(\bmW)\}}{1-p_s(\bmW)}\\
&\qquad\qquad+g'(\widehat\mu_1)\left\{\widehat m_1(\bmW)-\widehat\mu_1\right\}-g'(\widehat\mu_0)\left\{\widehat m_0(\bmW)-\widehat\mu_0\right\}\bigg],\qquad s=1,2.
\end{aligned}
\end{equation*}
Based on these observations, we propose to estimate $\delta$ using $\widehat\delta^{\text{aipw}}(\widehat c_1(\cdot;\widehat\eta),\widehat c_2(\cdot;\widehat\eta),\widehat\eta)$. Its asymptotic property is given in the next theorem, where we abbreviate $\widehat c_s=\widehat c_s(\cdot;\widehat\eta)$ and $c_s^*=c_s^*(\cdot;\eta_{\text{opt}}^*)$, $s=1,2$.

\begin{thm}\label{cdr.thm7}
Assume the conditions in Theorem \ref{cdr.thm5} hold. For any random triplet $(\widetilde c_1,\widetilde c_2,\widetilde\eta)$ that element-wise converges in probability to a fixed triplet $(c_1,c_2,\eta)$, we have, as $n_1\rightarrow\infty$,
\begin{equation*}\label{cdr.asyv.est}
\sqrt{n_1}\left\{\widehat\delta^{\text{aipw}}(\widetilde c_1,\widetilde c_2,\widetilde\eta)-\delta\right\}\stackrel{d}{\longrightarrow} N\left(0, \sigma_{\textsc{cdr}}^2(c_1,c_2,\eta)\right);
\end{equation*}
that is, $\widehat\delta^{\text{aipw}}(\widetilde c_1,\widetilde c_2,\widetilde\eta)$ is asymptotically equivalent to $\widehat\delta^{\text{aipw}}(c_1,c_2,\eta)$. In particular, for $(\widehat c_1,\widehat c_2,\widehat\eta)$ defined in the preceding paragraph, $\widehat\delta^{\text{aipw}}(\widehat c_1,\widehat c_2,\widehat\eta)$ is asymptotically equivalent to $\widehat\delta^{\text{aipw}}(c_1^*,c_2^*,\eta_{\text{opt}}^*)$ and attains the smallest asymptotic variance among $\{\widehat\delta^{\text{aipw}}(\widehat c_1(\cdot;\eta),\widehat c_2(\cdot;\eta),\eta),\eta\in[0,1]\}$. If $(m_1^*,m_0^*)=(m_1,m_0)$, then $\widehat\delta^{\text{aipw}}(\widehat c_1,\widehat c_2,\widehat\eta)$ attains the smallest asymptotic variance among all estimators of the form $\widehat\delta^{\text{aipw}}(c_1,c_2,\eta)$.
\end{thm}

The asymptotic variance of $\widehat\delta^{\text{aipw}}(\widehat c_1,\widehat c_2,\widehat\eta)$, $\sigma_{\textsc{cdr}}^2(c_1^*,c_2^*,\eta_{\text{opt}}^*)$, is consistently estimated by
\begin{equation*}
\begin{aligned}
&\widehat\var_1\bigg[\frac{\widehat{\eta} g'(\widehat\mu_1)A_{1}(Y_{1}-\widehat\mu_1)}{p_1(\bmW)}-\frac{\widehat{\eta} g'(\widehat\mu_0)(1-A_{1})(Y_{1}-\widehat\mu_0)}{1-p_1(\bmW)}-\{A_{1}-p_1(\bmW)\}\widehat{c}_1(\bmW)\bigg]\\
&+\frac{n_1}{n_2}\widehat\var_2\bigg[\frac{(1-\widehat\eta)g'(\widehat\mu_1)A_{2}(Y_{2}-\widehat\mu_1)}{p_2(\bmW)}-\frac{(1-\widehat\eta)g'(\widehat\mu_0)(1-A_{2})(Y_{2}-\widehat\mu_0)}{1-p_2(\bmW)}-\{A_{2}-p_2(\bmW)\}\widehat{c}_2(\bmW)\bigg].
\end{aligned}
\end{equation*}

\section*{Appendix B: Extension to Multiple Stages}\label{mult.stg}

We now extend the two-stage adaptive CDR design and estimation method in Appendix A to an arbitrary (specified) number of stages. Because CDR is strictly more general than CIR, the multi-stage CDR design described here can be reduced to a multi-stage CIR design by restricting the randomization mechanism to CIR at each stage, and the related estimation method remains applicable with propensity scores replaced by constants.

Let us consider an adaptive CDR trial with $k\ge2$ stages and pre-specified sample sizes $n_s$, $s=1,\dots,k$. The trial data consist of $(\bmD_1,\dots,\bmD_k)$, where $\bmD_s$ denotes the patient-level data from stage $s$: $\{\bmO_{si}=(\bmW_{si},A_{si},Y_{si}),i=1,\dots,n_s\}$. The first two stages are exactly as described in Section 5. If $k>2$, a second interim analysis is conducted after stage 2 to re-estimate $(\mu_1,\mu_0)$, $(v_1,v_0)$ and $p_{\text{opt}}$ on the basis of $(\bmD_1,\bmD_2)$. The updated estimate of $p_{\text{opt}}$, say $p_3$, is then used as the propensity score function for treatment assignment in stage 3. This process is iterated until all $k$ stages are completed. For each $s\ge2$, upon conditioning on $p_s$, the $\bmO_{si}$'s are independent across $i$ and identically distributed as $\bmO_s=(\bmW,A_s,Y_s)$, where $\pr\{A_s=1|\bmW,Y(1),Y(0)\}=\pr(A_s=1|\bmW)=p_s(\bmW)$ and $Y_s=Y(A_s)=A_sY(1)+(1-A_s)Y(0)$.

The treatment effect estimation problem here resembles the one considered in Appendix A; therefore, we will omit theoretical details and proofs and focus on describing the estimation method and key results. For each $a\in\{0,1\}$, let $\widehat\mu_a^{\text{ipw}(s)}$ be defined as in  Appendix A for $s=1,\dots,k$, and let $\widehat\mu_a^{\text{ipw}}(\bmeta)=\sum_{s=1}^k\eta_s\widehat\mu_a^{\text{ipw}(s)}$, where $\bmeta=(\eta_1,\dots,\eta_k)$ is a collection of normalized weights satisfying $\eta_s\ge0$ $\forall s$ and $\sum_{s=1}^k\eta_s=1$. A generic AIPW estimator of $\delta$ is given by
$$
\widehat\delta^{\text{aipw}}(\bmc,\bmeta)=\widehat\delta^{\text{ipw}}(\bmeta)-\sum_{s=1}^k\widehat\epn_s\left[\{A_{s}-p_s(\bmW)\}c_s(\bmW)\right],
$$
where $\widehat\delta^{\text{ipw}}(\bmeta)=g(\widehat\mu_1^{\text{ipw}}(\bmeta))-g(\widehat\mu_0^{\text{ipw}}(\bmeta))$ and $\bmc=(c_1,\dots,c_k)$ is a collection of real-valued functions of $\bmW$ such that $\epn\{c_s(\bmW)^2\}<\infty$, $s=1,\dots,k$. For each $s\in\{2,\dots,k\}$, we assume that, as $n_1\rightarrow\infty$, $n_1/n_s$ converges to some $\lambda_s\in(0,\infty)$ and $p_s$ converges to a limit function $p_s^*$ in the sense described in Theorem S1. (For notational convenience, we define $\lambda_1=1$ and $p_1^*=p_1$.) Then, for any fixed $(\bmc,\bmeta)$, $\sqrt{n_1}\{\widehat\delta^{\text{aipw}}(\bmc,\bmeta)-\delta\}$ converges to a zero-mean normal distribution whose variance is denoted by $\sigma_{\textsc{cdr}}^2(\bmc,\bmeta)$.

For any given $\bmeta$, the asymptotic variance $\sigma_{\textsc{cdr}}^2(\bmc,\bmeta)$ is minimized by setting $\bmc=\bmc_{\text{opt}}(\cdot;\bmeta)$, where $\bmc_{\text{opt}}=(c_{1,\text{opt}},\dots,c_{k,\text{opt}})$ with
$$
c_{s,\text{opt}}(\bmW;\bmeta)=\eta_s\left[\frac{g'(\mu_1)\{m_1(\bmW)-\mu_1\}}{p_s^*(\bmW)}+
\frac{g'(\mu_0)\{m_0(\bmW)-\mu_0\}}{1-p_s^*(\bmW)}\right],\qquad s=1,\dots,k.
$$
The optimal choice $\bmc_{\text{opt}}$ is estimated by $\widehat\bmc=(\widehat c_1,\dots,\widehat c_k)$, where
$$
\widehat c_{s}(\bmW;\bmeta)=\eta_s\left[\frac{g'(\widehat\mu_1)\{\widehat m_1(\bmW)-\widehat\mu_1\}}{p_s(\bmW)}+
\frac{g'(\widehat\mu_0)\{\widehat m_0(\bmW)-\widehat\mu_0\}}{1-p_s(\bmW)}\right],\qquad s=1,\dots,k.
$$
With $(\widehat\mu_1,\widehat\mu_0)$ converging to $(\mu_1,\mu_0)$ and $(\widehat m_1,\widehat m_0)$ to $(m_1^*,m_0^*)$, $\widehat\bmc$ converges to $\bmc^*=(c_1^*,\dots,c_k^*)$, where
$$
c_{s}^*(\bmW;\bmeta)=\eta_s\left[\frac{g'(\mu_1)\{m_1^*(\bmW)-\mu_1\}}{p_s^*(\bmW)}+
\frac{g'(\mu_0)\{m_0^*(\bmW)-\mu_0\}}{1-p_s^*(\bmW)}\right],\qquad s=1,\dots,k.
$$
It can be shown that $\sigma_{\textsc{cdr}}^2(\bmc^*(\cdot;\bmeta),\bmeta)$ is minimized by setting $\bmeta$ equal to $\bmeta_{\text{opt}}^*=(\eta_{1,\text{opt}}^*,\dots,\eta_{k,\text{opt}}^*)$, where, for each $s\in\{1,\dots,k\}$,
\begin{equation*}
\begin{aligned}
\eta_{s,\text{opt}}^*&=\left(\lambda_s\sigma_{s,\textsc{cdr}}^{*2}\right)^{-1}\left/\sum_{j=1}^k\left(\lambda_j\sigma_{j,\textsc{cdr}}^{*2}\right)^{-1}\right.,\\
\sigma_{s,\textsc{cdr}}^{*2}&=\var\bigg[\frac{g'(\mu_1)A_s^*\{Y_s^*-m_1^*(\bmW)\}}{p_s^*(\bmW)}-\frac{g'(\mu_0)(1-A_s^*)\{Y_s^*-m_0^*(\bmW)\}}{1-p_s^*(\bmW)}\\
&\qquad\quad+g'(\mu_1)\left\{m_1^*(\bmW)-\mu_1\right\}-g'(\mu_0)\left\{m_0^*(\bmW)-\mu_0\right\}\bigg],\\
\end{aligned}
\end{equation*}
and $\bmO_s^*=(\bmW,A_s^*,Y_s^*)$ is similar to $\bmO_s$ with $p_s^*$ replacing $p_s$. For any $(m_1^*,m_0^*)$, $\bmeta_{\text{opt}}^*$ is consistently estimated by $\widehat\bmeta=(\widehat\eta_1,\dots,\widehat\eta_k)$, where, for each $s\in\{1,\dots,k\}$,
\begin{equation*}
\begin{aligned}
\widehat\eta_{s}&=n_s\widehat\sigma_{s,\textsc{cdr}}^{-2}\left/\sum_{j=1}^kn_j\widehat\sigma_{j,\textsc{cdr}}^{-2}\right.,\\
\widehat\sigma_{s,\textsc{cdr}}^{2}&=\widehat\var_s\bigg[\frac{g'(\widehat\mu_1)A_s\{Y_s-\widehat m_1(\bmW)\}}{p_s(\bmW)}-\frac{g'(\widehat\mu_0)(1-A_s)\{Y_s-\widehat m_0(\bmW)\}}{1-p_s(\bmW)}\\
&\qquad\quad+g'(\widehat\mu_1)\left\{\widehat m_1(\bmW)-\widehat\mu_1\right\}-g'(\widehat\mu_0)\left\{\widehat m_0(\bmW)-\widehat\mu_0\right\}\bigg].\\
\end{aligned}
\end{equation*}

Based on the above optimality results, we propose to estimate $\delta$ using $\widehat\delta^{\text{aipw}}(\widehat\bmc(\cdot;\widehat\bmeta),\widehat\bmeta)$, which is asymptotically equivalent to $\widehat\delta^{\text{aipw}}(\bmc^*(\cdot;\bmeta_{\text{opt}}^*),\bmeta_{\text{opt}}^*)$. This estimator attains the smallest asymptotic variance among all estimators of the form $\widehat\delta^{\text{aipw}}(\widehat\bmc(\cdot;\bmeta),\bmeta)$ and, if $(m_1^*,m_0^*)=(m_1,m_0)$, among all AIPW estimators considered here. Its asymptotic variance is consistently estimated by
$$
\sum_{s=1}^k\frac{n_1}{n_s}\widehat\var_s\bigg[\frac{\widehat\eta_sg'(\widehat\mu_1)A_{s}(Y_{s}-\widehat\mu_1)}{p_s(\bmW)}-\frac{\widehat\eta_sg'(\widehat\mu_0)(1-A_{s})(Y_{s}-\widehat\mu_0)}{1-p_s(\bmW)}-\{A_{s}-p_s(\bmW)\}\widehat{c}_s(\bmW)\bigg].
$$

\section*{Appendix C: Proofs}\label{proofs}

\subsection*{Proof of Theorem 1}

Using the Taylor expansions of the function $g(\cdot)$ at $\mu_1$ and $\mu_0$, we have
\begin{equation*}
\begin{aligned}
&\widehat\delta^{\text{aug}}(b_1,b_2,\theta)-\delta\\
&=\left\{g(\widehat\mu_1)-g(\mu_1)\right\}-\left\{g(\widehat\mu_0)-g(\mu_0)\right\}
-\frac{1}{n_1}\sum_{i=1}^{n_1}(A_{1i}-\pi_1)b_1(\bmW_{1i})-\frac{1}{n_2}\sum_{j=1}^{n_2}(A_{2j}-\pi_2)b_2(\bmW_{2j})\\
&=\theta g'(\mu_1)\left(\widehat\mu_{11}-\mu_1\right)-\theta g'(\mu_0)\left(\widehat\mu_{10}-\mu_0\right)-\frac{1}{n_1}\sum_{i=1}^{n_1}(A_{1i}-\pi_1)b_1(\bmW_{1i})\\
&\quad+(1-\theta)g'(\mu_1)\left(\widehat\mu_{21}-\mu_1\right)-(1-\theta)g'(\mu_0)\left(\widehat\mu_{20}-\mu_0\right)-\frac{1}{n_2}\sum_{j=1}^{n_2}(A_{2j}-\pi_2)b_2(\bmW_{2j})\\
&\quad+O_p\left\{(\widehat\mu_1-\mu_1)^2\right\}+O_p\left\{(\widehat\mu_0-\mu_0)^2\right\}\\
&=Z_1+Z_2+O_p\left\{(\widehat\mu_1-\mu_1)^2\right\}+O_p\left\{(\widehat\mu_0-\mu_0)^2\right\},
\end{aligned}
\end{equation*}
where
\begin{equation*}
\begin{aligned}
Z_1&=\theta g'(\mu_1)\left(\widehat\mu_{11}-\mu_1\right)-\theta g'(\mu_0)\left(\widehat\mu_{10}-\mu_0\right)-\frac{1}{n_1}\sum_{i=1}^{n_1}(A_{1i}-\pi_1)b_1(\bmW_{1i}),\\
Z_2&=(1-\theta)g'(\mu_1)\left(\widehat\mu_{21}-\mu_1\right)-(1-\theta)g'(\mu_0)\left(\widehat\mu_{20}-\mu_0\right)-\frac{1}{n_2}\sum_{j=1}^{n_2}(A_{2j}-\pi_2)b_2(\bmW_{2j}).
\end{aligned}
\end{equation*}
The stage-specific treatment group averages for $\mu_1$ and $\mu_0$ are given by
\begin{equation*}
\begin{aligned}
\widehat\mu_{11}&=\frac{\sum_{i=1}^{n_1}I(A_{1i}=1)Y_{1i}}{\sum_{i=1}^{n_1}I(A_{1i}=1)}, \qquad
\widehat\mu_{10}=\frac{\sum_{i=1}^{n_1}I(A_{1i}=0)Y_{1i}}{\sum_{i=1}^{n_1}I(A_{1i}=0)},\\
\widehat\mu_{21}&=\frac{\sum_{j=1}^{n_2}I(A_{2j}=1)Y_{2j}}{\sum_{j=1}^{n_2}I(A_{2j}=1)},\qquad \widehat\mu_{20}=\frac{\sum_{j=1}^{n_2}I(A_{2j}=0)Y_{2j}}{\sum_{j=1}^{n_2}I(A_{2j}=0)}.
\end{aligned}
\end{equation*}
The law of large numbers ensures that, as $n_1\rightarrow\infty$,
\begin{equation*}
\frac{1}{n_1}\sum_{i=1}^{n_1}\frac{I(A_{1i}=1)}{\pi_1}=1+o_p(1)\quad\text{and}\quad\frac{1}{n_1}\sum_{i=1}^{n_1}\frac{I(A_{1i}=0)}{1-\pi_1}=1+o_p(1).
\end{equation*}
Then we have
\begin{equation*}
\begin{aligned}
Z_1&=\theta g'(\mu_1)\frac{\sum_{i=1}^{n_1}I(A_{1i}=1)(Y_{1i}-\mu_1)}{\sum_{i=1}^{n_1}I(A_{1i}=1)}-\theta g'(\mu_0)\frac{\sum_{i=1}^{n_1}I(A_{1i}=0)(Y_{1i}-\mu_0)}{\sum_{i=1}^{n_1}I(A_{1i}=0)}\hspace{2.7cm}\\
\end{aligned}
\end{equation*}
\begin{equation*}
\begin{aligned}
&\quad-\frac{1}{n_1}\sum_{i=1}^{n_1}(A_{1i}-\pi_1)b_1(\bmW_{1i})\\
&=\theta g'(\mu_1)\frac{\sum_{i=1}^{n_1}A_{1i}(Y_{1i}-\mu_1)}{n_1\pi_1}\{1+o_p(1)\}-\theta g'(\mu_0)\frac{\sum_{i=1}^{n_1}(1-A_{1i})(Y_{1i}-\mu_0)}{n_1(1-\pi_1)}\{1+o_p(1)\}\\
&\quad-\frac{1}{n_1}\sum_{i=1}^{n_1}(A_{1i}-\pi_1)b_1(\bmW_{1i})\\
&=\frac{1}{n_1}\sum_{i=1}^{n_1}\left\{\frac{\theta g'(\mu_1)A_{1i}(Y_{1i}-\mu_1)}{\pi_1}-\frac{\theta g'(\mu_0)(1-A_{1i})(Y_{1i}-\mu_0)}{1-\pi_1}-(A_{1i}-\pi_1)b_1(\bmW_{1i})\right\}\\
&\qquad\qquad\times\{1+o_p(1)\}\\
&=\frac{1}{n_1}\sum\limits_{i=1}^{n_1}\psi_1^{\text{aug}}(\bmO_{1i})\{1+o_p(1)\},
\end{aligned}
\end{equation*}
where
$$\psi_1^{\text{aug}}(\bmO_1)=\frac{\theta g'(\mu_1)A_1(Y_1-\mu_1)}{\pi_1}-\frac{\theta g'(\mu_0)(1-A_1)(Y_1-\mu_0)}{1-\pi_1}-(A_1-\pi_1)b_1(\bmW).$$
The central limit theorem ensures that, as $n_1\rightarrow\infty$,
\begin{equation}\label{stage1AsyDis.cir}\tag{A.1}
\sqrt{n_1}Z_1\stackrel{d}{\longrightarrow} N\left(0, \var\left\{\psi_1^{\text{aug}}(\bmO_1)\right\}\right).
\end{equation}

Conditioning on the stage 1 data $\bmD_1$, $n_2$ and $\pi_2$ are known. Then we have
\begin{equation*}
\frac{1}{n_2}\sum_{j=1}^{n_2}\frac{I(A_{2j}=1)}{\pi_2}=1+o_p(1)\quad\text{and}\quad\frac{1}{n_2}\sum_{j=1}^{n_2}\frac{I(A_{2j}=0)}{1-\pi_2}=1+o_p(1), \quad\text{as}\quad n_2\rightarrow\infty.
\end{equation*}
Substituting these into the expression of $Z_2$ yields
\begin{equation*}
\begin{aligned}
Z_2&=(1-\theta) g'(\mu_1)\frac{\sum_{j=1}^{n_2}A_{2j}(Y_{2j}-\mu_1)}{n_2\pi_2}\{1+o_p(1)\}\\
&\quad-(1-\theta) g'(\mu_0)\frac{\sum_{j=1}^{n_2}(1-A_{2j})(Y_{2j}-\mu_0)}{n_2(1-\pi_2)}\{1+o_p(1)\}-\frac{1}{n_2}\sum_{j=1}^{n_2}(A_{2j}-\pi_2)b_2(\bmW_{2j})\\
&=\frac{1}{n_2}\sum_{j=1}^{n_2}\Big\{\frac{(1-\theta)g'(\mu_1)A_{2j}(Y_{2j}-\mu_1)}{\pi_2}-\frac{(1-\theta)g'(\mu_0)(1-A_{2j})(Y_{2j}-\mu_0)}{1-\pi_2}\\
&\qquad\qquad\quad-(A_{2j}-\pi_2)b_2(\bmW_{2j})\Big\}\{1+o_p(1)\}\\
&=\frac{1}{n_2}\sum_{j=1}^{n_2}\psi_2^{\text{aug}}(\bmO_{2j})\{1+o_p(1)\},
\end{aligned}
\end{equation*}
where
$$\psi_2^{\text{aug}}(\bmO_2)=\frac{(1-\theta)g'(\mu_1)A_{2}(Y_{2}-\mu_1)}{\pi_2}-\frac{(1-\theta)g'(\mu_0)(1-A_{2})(Y_{2}-\mu_0)}{1-\pi_2}-(A_{2}-\pi_2)b_2(\bmW).$$
Applying the central limit theorem yields
\begin{equation}\label{stage2AsyDis.cir}\tag{A.2}
\sqrt{n_2}Z_2 \mid \bmD_1\stackrel{d}{\longrightarrow} N(0, \var\{\psi_2^{\text{aug}}(\bmO_2)\}), \quad\text{as } n_2\rightarrow\infty.
\end{equation}

Moreover, we observe that
\begin{equation*}
\begin{aligned}
&\pr(\sqrt{n_1}Z_1\leq z_1, \sqrt{n_2}Z_2\leq z_2)\\
&=\pr(\sqrt{n_1}Z_1\leq z_1)\pr(\sqrt{n_2}Z_2\leq z_2\,|\,\sqrt{n_1}Z_1\leq z_1)\\
&=\pr(\sqrt{n_1}Z_1\leq z_1)\epn\left\{I(\sqrt{n_2}Z_2\leq z_2)\;|\;\sqrt{n_1}Z_1\leq z_1\right\}\\
&=\pr(\sqrt{n_1}Z_1\leq z_1)\epn\left[\epn\left\{I(\sqrt{n_2}Z_2\leq z_2) | \bmD_1, \sqrt{n_1}Z_1\leq z_1\right\} | \sqrt{n_1}Z_1\leq z_1\right]\\
&=\pr(\sqrt{n_1}Z_1\leq z_1)\epn\left\{\pr(\sqrt{n_2}Z_2\leq z_2 | \bmD_1) | \sqrt{n_1}Z_1\leq z_1\right\}
\end{aligned}
\end{equation*}
By \eqref{stage2AsyDis.cir}, we have $\pr(\sqrt{n_2}Z_2\leq z_2|\bmD_1)\rightarrow\Phi(z_2/\sqrt{\var\{\psi_2^{\text{aug}}(\bmO_2)\}})$ as $n_2\rightarrow\infty$, where $\Phi(\cdot)$ is the cumulative distribution function of the standard normal distribution. Further letting $n_1\rightarrow\infty$, we can obtain that $\pr(\sqrt{n_1}Z_1\leq z_1)\rightarrow\Phi(z_1/\sqrt{\var\{\psi_1^{\text{aug}}(\bmO_1)\}})$ and $\Phi(z_2/\sqrt{\var\{\psi_2^{\text{aug}}(\bmO_2)\}})\rightarrow\Phi(z_2/\sqrt{\var\{\psi_2^{\text{aug}}(\bmO_2^*)\}})$, where
\begin{equation*}
\begin{aligned}
\psi_2^{\text{aug}}(\bmO_2^*)&=\frac{(1-\theta)g'(\mu_1)A_{2}^*(Y_{2}^*-\mu_1)}{\pi_2^*}-\frac{(1-\theta)g'(\mu_0)(1-A_{2}^*)(Y_{2}^*-\mu_0)}{1-\pi_2^*}-(A_{2}^*-\pi_2^*)b_2(\bmW)
\end{aligned}
\end{equation*}
and $\bmO_2^*=(\bmW,A_2^*,Y_2^*)$ is similar to $\bmO_2$ with $\pi_2^*$ replacing $\pi_2$,
$\pr\{A_2^*=1|\bmW,Y(1),Y(0)\}=\pr(A_2^*=1)=\pi_2^*$, and
$Y_2^*=Y(A_2^*)=A_2^*Y(1)+(1-A_2^*)Y(0)$. Since $n_1/n_2\rightarrow\lambda$ as $n_1\rightarrow\infty$, then we have
\begin{equation*}
\pr(\sqrt{n_1}Z_1\leq z_1, \sqrt{n_2}Z_2\leq z_2)\rightarrow\Phi\left(\frac{z_1}{\sqrt{\var\{\psi_1^{\text{aug}}(\bmO_1)\}}}\right)\Phi\left(\frac{z_2}{\sqrt{\var\{\psi_2^{\text{aug}}(\bmO_2^*)\}}}\right),
\end{equation*}
as $n_1\rightarrow\infty$. This implies that, as $n_1\rightarrow\infty$,
\begin{equation*}
\sqrt{n_1}\begin{pmatrix}Z_1\\Z_2\end{pmatrix}\stackrel{d}{\longrightarrow}N\left(\begin{pmatrix}0\\0\end{pmatrix}, \begin{pmatrix}\var\{\psi_1^{\text{aug}}(\bmO_1)\}&0\\0&\lambda\var\{\psi_2^{\text{aug}}(\bmO_2^*)\}\end{pmatrix}\right).
\end{equation*}
Applying the delta method yields
$$\sqrt{n_1}(Z_1+Z_2)\stackrel{d}{\longrightarrow}N\left(0, \var\{\psi_1^{\text{aug}}(\bmO_1)\}+\lambda\var\{\psi_2^{\text{aug}}(\bmO_2^*)\}\right).$$ The Slutsky's theorem ensures that
\begin{equation*}
\sqrt{n_1}\left\{\widehat\delta^{\text{aug}}(b_1,b_2,\theta)-\delta\right\}\stackrel{d}{\longrightarrow} N\left(0, \var\{\psi_1^{\text{aug}}(\bmO_1)\}+\lambda\var\{\psi_2^{\text{aug}}(\bmO_2^*)\}\right).
\end{equation*}

\subsection*{Derivation of $(b_{1,\text{opt}},b_{2,\text{opt}})$}

Note that $\var\{\psi_1^{\text{aug}}(\bmO_1)\}$ (rsp. $\var\{\psi_2^{\text{aug}}(\bmO_2^*)\}$) is a function of $b_1$ (rsp. $b_2$) only. Minimizing the asymptotic variance $\sigma_{\textsc{cir}}^2(b_1,b_2,\theta)=\var\{\psi_1^{\text{aug}}(\bmO_1)\}+\lambda\var\{\psi_2^{\text{aug}}(\bmO_2^*)\}$ with respect to $b_1$ (rsp. $b_2$) is equivalent to minimizing $\var\{\psi_1^{\text{aug}}(\bmO_1)\}$ (rsp. $\var\{\psi_2^{\text{aug}}(\bmO_2^*)\}$) with respect to $b_1$ (rsp. $b_2$). Simple algebra leads to
\begin{equation*}
\begin{aligned}
\var\{\psi_1^{\text{aug}}(\bmO_1)\}&=\epn\left(\epn\left[\left\{\frac{\theta g'(\mu_1)A_{1}(Y_{1}-\mu_1)}{\pi_1}-\frac{\theta g'(\mu_0)(1-A_{1})(Y_{1}-\mu_0)}{1-\pi_1}\right\}^2\bigg| \bmW\right]\right)\\
&\quad-2\epn\left[\epn\left\{\theta(A_1-\pi_1)g'(\mu_1)\frac{A_1(Y_1-\mu_1)}{\pi_1} \bigg|\bmW\right\}b_1(\bmW)\right]\\
&\quad+2\epn\left[\epn\left\{\theta(A_1-\pi_1)g'(\mu_0)\frac{(1-A_1)(Y_1-\mu_0)}{1-\pi_1} \bigg|\bmW\right\}b_1(\bmW)\right]\\
&\quad+\epn\left[\epn\left\{(A_1-\pi_1)^2|\bmW\right\}b_1(\bmW)^2\right]
\end{aligned}
\end{equation*}
The minimizer of $\var\{\psi_1^{\text{aug}}(\bmO_1)\}$ with respect to $b_1$ is easily found to be
\begin{equation*}
\begin{aligned}
b_{1,\text{opt}}(\bmW;\theta)&=\frac{2\epn\left[\theta(A_1-\pi_1)\left\{g'(\mu_1)A_1(Y_1-\mu_1)/\pi_1-g'(\mu_0)(1-A_1)(Y_1-\mu_0)/(1-\pi_1) \right\}|\bmW\right]}{2\epn\left\{(A_1-\pi_1)^2|\bmW\right\}}\\
&=\theta\left[\frac{g'(\mu_1)\{m_1(\bmW)-\mu_1\}}{\pi_1}+
\frac{g'(\mu_0)\{m_0(\bmW)-\mu_0\}}{1-\pi_1}\right],
\end{aligned}
\end{equation*}
where $m_a(\bmW)=\epn\{Y(a)|\bmW\}=\epn(Y|\bmW,A=a)$, $a=0,1$. Similarly, the minimizer of $\var\{\psi_2^{\text{aug}}(\bmO_2^*)\}$ with respect to $b_2$ can be obtained as
\begin{equation*}
\begin{aligned}
b_{2,\text{opt}}(\bmW;\theta)&=\frac{2\epn\left[(1-\theta)(A_2-\pi_2^*)\left\{g'(\mu_1)A_2(Y_2-\mu_1)/\pi_2^* \right\}|\bmW\right]}{2\epn\left\{(A_2-\pi_2^*)^2|\bmW\right\}}\\
&\quad-\frac{2\epn\left[(1-\theta)(A_2-\pi_2^*)\left\{g'(\mu_0)(1-A_2)(Y_2-\mu_0)/(1-\pi_2^*) \right\}|\bmW\right]}{2\epn\left\{(A_2-\pi_2^*)^2|\bmW\right\}}\\
&=(1-\theta)\left[\frac{g'(\mu_1)\{m_1(\bmW)-\mu_1\}}{\pi_2^*}+
\frac{g'(\mu_0)\{m_0(\bmW)-\mu_0\}}{1-\pi_2^*}\right].
\end{aligned}
\end{equation*}

\subsection*{Derivation of $\theta_{\text{opt}}^*$}

Substituting $(b_1^*,b_2^*)$ into $\sigma_{\textsc{cir}}^2(b_1,b_2,\theta)$ yields
\begin{equation*}
\begin{aligned}
\sigma_{\textsc{cir}}^2(b_1^*,b_2^*,\theta)&=\theta^2\sigma_{1,\textsc{cir}}^{*2}+\lambda(1-\theta)^2\sigma_{2,\textsc{cir}}^{*2},
\end{aligned}
\end{equation*}
where
\begin{equation*}
\begin{aligned}
\sigma_{1,\textsc{cir}}^{*2}&=\var\bigg[\frac{g'(\mu_1)A_1\{Y_1-m_1^*(\bmW)\}}{\pi_1}-\frac{g'(\mu_0)(1-A_1)\{Y_1-m_0^*(\bmW)\}}{1-\pi_1}\\
&\qquad\quad+g'(\mu_1)\left\{m_1^*(\bmW)-\mu_1\right\}-g'(\mu_0)\left\{m_0^*(\bmW)-\mu_0\right\}\bigg],\\
\sigma_{2,\textsc{cir}}^{*2}&=\var\bigg[\frac{g'(\mu_1)A_2^*\{Y_2^*-m_1^*(\bmW)\}}{\pi_2^*}-\frac{g'(\mu_0)(1-A_2^*)\{Y_2^*-m_0^*(\bmW)\}}{1-\pi_2^*}\\
&\qquad\quad+g'(\mu_1)\left\{m_1^*(\bmW)-\mu_1\right\}-g'(\mu_0)\left\{m_0^*(\bmW)-\mu_0\right\}\bigg].\\
\end{aligned}
\end{equation*}
The minimizer of $\sigma_{\textsc{cir}}^2(b_1^*,b_2^*,\theta)$ with respect to $\theta$ is then equal to
\begin{equation*}
\theta_{\text{opt}}^*=\frac{\lambda\sigma_{2,\textsc{cir}}^{*2}}{\sigma_{1,\textsc{cir}}^{*2}+\lambda\sigma_{2,\textsc{cir}}^{*2}}.
\end{equation*}

\subsection*{Convergence of $(\widehat m_1,\widehat m_0)$}

Suppose we specify a regression model for the conditional mean, i.e., $\epn(Y_1|A_1=a,\bmW)=m_a(\bmW; \xi)$. The parameter $\xi$ is usually estimated by solving an estimating equation:
$$\sum_{i=1}^{n_1}h(\bmO_{1i}; \xi)+\sum_{i=1}^{n_2}h(\bmO_{2i}; \xi)=0,$$
where $h$ is an estimating function of the same dimension as $\xi$. Let $\widehat{\xi}$ denote the resulting estimator, which may be the maximum likelihood estimator for logistic regression (used in our simulation study). It can be argued as in the proof of Theorem 1 that $n^{-1}\sum_{i=1}^{n_1}h(\bmO_{1i}; \xi)+n^{-1}\sum_{i=1}^{n_2}h(\bmO_{2i}; \xi)$ converges in probability to $$\lambda/(1+\lambda)\epn\{h(\bmO_1;\xi)\}+1/(1+\lambda)\epn\{h(\bmO_2^*;\xi)\}.$$ Under certain regularity conditions (e.g., if $h$ belongs to a Glivenko-Cantelli class; see van der Vaart (1998, Chapter 19)), the convergence in probability is uniform. Let $\xi^*$ be the solution to the equation $\lambda/(1+\lambda)\epn\{h(\bmO_1;\xi)\}+1/(1+\lambda)\epn\{h(\bmO_2^*;\xi)\}=0$. By Theorem 5.9 of van der Vaart (1998), $\widehat{\xi}$ converges in probability to $\xi^*$, and therefore $\widehat{m}_a=m_a(\cdot;\widehat\xi)$ converges to $m_a^*=m_a(\cdot;\xi^*)$.

\subsection*{Consistency of $\widehat{\theta}$}

The consistency of $\widehat{\theta}$ follows from the consistency of the variance estimators $\widehat\sigma_{1,\textsc{cir}}^2$ and $\widehat\sigma_{2,\textsc{cir}}^2$. First, since the stage 1 data $\bmD_1$ are independent and identically distributed, the sample variance estimator $\widehat\sigma_{1,\textsc{cir}}^2$ converges in probability to $\sigma_{1,\textsc{cir}}^{*2}$ by the law of large numbers. For Stage 2, note that for any $\varepsilon>0$,
$$\pr(|\widehat\sigma_{2,\textsc{cir}}^2-\sigma_{2,\textsc{cir}}^{*2}|>\varepsilon)\leq\pr(|\widehat\sigma_{2,\textsc{cir}}^2-\sigma_{2,\textsc{cir}}^{2}|>\varepsilon/2)+
\pr(|\sigma_{2,\textsc{cir}}^{2}-\sigma_{2,\textsc{cir}}^{*2}|>\varepsilon/2),$$
where $\sigma_{2,\textsc{cir}}^{2}$ is defined analogously to $\sigma_{2,\textsc{cir}}^{*2}$ as
\begin{equation*}
\begin{aligned}
\sigma_{2,\textsc{cir}}^{2}&=\var\bigg[\frac{g'(\mu_1)A_2\{Y_2-m_1^*(\bmW)\}}{\pi_2}-\frac{g'(\mu_0)(1-A_2)\{Y_2-m_0^*(\bmW)\}}{1-\pi_2}\\
&\qquad\quad+g'(\mu_1)\left\{m_1^*(\bmW)-\mu_1\right\}-g'(\mu_0)\left\{m_0^*(\bmW)-\mu_0\right\}\bigg].\\
\end{aligned}
\end{equation*}
but with $\pr\{A_2=1|\bmW,Y(1),Y(0)\}=\pr(A_2=1)=\pi_2$.

For the first term, by the law of total probability, $\pr(|\widehat\sigma_{2,\textsc{cir}}^2-\sigma_{2,\textsc{cir}}^{2}|>\varepsilon/2)=\epn\{\pr(|\widehat\sigma_{2,\textsc{cir}}^2-\sigma_{2,\textsc{cir}}^{2}|>\varepsilon/2 \mid \pi_2)\}$. The inner probability $\pr(|\widehat\sigma_{2,\textsc{cir}}^2-\sigma_{2,\textsc{cir}}^{2}|>\varepsilon/2 \mid \pi_2)$ converges to zero as $n_2\rightarrow\infty$, since conditional on $\pi_2$, the stage 2 data $\bmD_2$ are independent and identically distributed. Therefore, by the bounded convergence theorem, the first term converges to zero. For the second term, note that $\sigma_{2,\textsc{cir}}^{2}$ and $\sigma_{2,\textsc{cir}}^{*2}$ are continuous functionals of $\pi_2$ and $\pi_2^*$, respectively. Under the assumption that $\pi_2$ converges in probability to $\pi_2^*$, the second term converges to zero as $n_1\to\infty$ by the continuous mapping theorem. Combining these results implies that $\widehat\sigma_{2,\textsc{cir}}^2$ converges in probability to $\sigma_{2,\textsc{cir}}^{*2}$. Consequently, by Slutsky's theorem, the estimator $\hat{\theta}$ is consistent for $\theta_{\text{opt}}^*$.

\subsection*{Proof of Theorem 2}

According to the proof of Theorem 1, we can write
\begin{equation*}
\begin{aligned}
&\widehat\delta^{\text{aug}}(\widetilde b_1,\widetilde b_2,\widetilde\theta)-\delta\\
&=\frac{1}{n_1}\sum_{i=1}^{n_1}\left\{\frac{\widetilde\theta g'(\mu_1)A_{1i}(Y_{1i}-\mu_1)}{\pi_1}-\frac{\widetilde\theta g'(\mu_0)(1-A_{1i})(Y_{1i}-\mu_0)}{1-\pi_1}-(A_{1i}-\pi_1)\widetilde{b}_1(\bmW_{1i})\right\}\\
&\quad+\frac{1}{n_2}\sum_{j=1}^{n_2}\bigg\{\frac{(1-\widetilde\theta)g'(\mu_1)A_{2j}(Y_{2j}-\mu_1)}{\pi_2}-\frac{(1-\widetilde\theta)g'(\mu_0)(1-A_{2j})(Y_{2j}-\mu_0)}{1-\pi_2}\\
&\qquad\qquad\qquad-(A_{2j}-\pi_2)\widetilde{b}_2(\bmW_{2j})\bigg\}+O_p\left\{(\widehat\mu_1-\mu_1)^2\right\}+O_p\left\{(\widehat\mu_0-\mu_0)^2\right\}\\
&=\frac{1}{n_1}\sum_{i=1}^{n_1}\left\{\frac{\theta g'(\mu_1)A_{1i}(Y_{1i}-\mu_1)}{\pi_1}-\frac{\theta g'(\mu_0)(1-A_{1i})(Y_{1i}-\mu_0)}{1-\pi_1}-(A_{1i}-\pi_1)b_1(\bmW_{1i})\right\}\\
&\quad+\frac{\widetilde{\theta}-\theta}{n_1}\sum\limits_{i=1}^{n_1}\left\{\frac{ g'(\mu_1)A_{1i}(Y_{1i}-\mu_1)}{\pi_1}-\frac{ g'(\mu_0)(1-A_{1i})(Y_{1i}-\mu_0)}{1-\pi_1}\right\}\\
&\quad-\frac{1}{n_1}\sum\limits_{i=1}^{n_1}(A_{1i}-\pi_1)\left\{\widetilde{b}_1(\bmW_{1i})-b_1(\bmW_{1i})\right\}\\
\end{aligned}
\end{equation*}
\begin{equation*}
\begin{aligned}
&\quad+\frac{1}{n_2}\sum_{j=1}^{n_2}\bigg\{\frac{(1-\theta)g'(\mu_1)A_{2j}(Y_{2j}-\mu_1)}{\pi_2}-\frac{(1-\theta)g'(\mu_0)(1-A_{2j})(Y_{2j}-\mu_0)}{1-\pi_2}\qquad\quad\\
&\qquad\qquad\qquad-(A_{2j}-\pi_2)b_2(\bmW_{2j})\bigg\}\\
&\quad+\frac{\theta-\widetilde{\theta}}{n_2}\sum\limits_{j=1}^{n_2}\bigg\{\frac{g'(\mu_1)A_{2j}(Y_{2j}-\mu_1)}{\pi_2}-\frac{g'(\mu_0)(1-A_{2j})(Y_{2j}-\mu_0)}{1-\pi_2}\bigg\}\\
&\quad-\frac{1}{n_2}\sum\limits_{j=1}^{n_2}(A_{2j}-\pi_2)\left\{\widetilde{b}_2(\bmW_{2j})-b_2(\bmW_{2j})\right\}+o_p(\widehat\mu_1-\mu_1)+o_p(\widehat\mu_0-\mu_0).
\end{aligned}
\end{equation*}
The condition $\widetilde{\theta}\rightarrow\theta$ as $n_1\rightarrow\infty$ ensures that
\begin{equation*}
\begin{aligned}
&\frac{\widetilde{\theta}-\theta}{n_1}\sum\limits_{i=1}^{n_1}\left\{\frac{ g'(\mu_1)A_{1i}(Y_{1i}-\mu_1)}{\pi_1}-\frac{ g'(\mu_0)(1-A_{1i})(Y_{1i}-\mu_0)}{1-\pi_1}\right\}\\
&=o_p\left[\frac{1}{n_1}\sum\limits_{i=1}^{n_1}\left\{\frac{ g'(\mu_1)A_{1i}(Y_{1i}-\mu_1)}{\pi_1}-\frac{ g'(\mu_0)(1-A_{1i})(Y_{1i}-\mu_0)}{1-\pi_1}\right\}\right]\\
&=o_p(1/\sqrt{n_1}).
\end{aligned}
\end{equation*}
Moreover, by the Chebyshev inequality \citep{S03}, for any real number $\varepsilon>0$, we have
\begin{equation*}
\begin{aligned}
&\pr\left[\frac{1}{\sqrt{n_1}}\sum\limits_{i=1}^{n_1}(A_{1i}-\pi_1)\left\{\widetilde{b}_1(\bmW_{1i})-b_1(\bmW_{1i})\right\}>\varepsilon\right]\\
&\leq\frac{\epn\left[\sum\limits_{i=1}^{n_1}(A_{1i}-\pi_1)\left\{\widetilde{b}_1(\bmW_{1i})-b_1(\bmW_{1i})\right\}\right]^2}{n_1\varepsilon^2}\\
&=\frac{\epn\left[(A_{1i}-\pi_1)^2\left\{\widetilde{b}_1(\bmW_{1i})-b_1(\bmW_{1i})\right\}^2\right]}{\varepsilon^2}\\
&\leq\frac{\epn\left\{\widetilde{b}_1(\bmW_{1i})-b_1(\bmW_{1i})\right\}^2}{\varepsilon^2}\\
&=o_p(1),
\end{aligned}
\end{equation*}
where the second inequality is due to $|A_{1i}-\pi_1|\leq1$ and the last equality is due to the condition $\|\widetilde{b}_1-b_1^*\|_2=[\epn\{\widetilde{b}_1-b_1^*\}^2]^{1/2}=o_p(1)$, so that
$$\frac{1} {\sqrt{n_1}}\sum\limits_{i=1}^{n_1}(A_{1i}-\pi_1)\left\{\widetilde{b}_1(\bmW_{1i})-b_1(\bmW_{1i})\right\}=o_p(1).$$

Conditioning on $\bmD_1$, $n_2$ and $\pi_2$ are known. Then we can similarly obtain that
\begin{equation*}
\begin{aligned}
\frac{\theta^*-\widetilde{\theta}}{n_2}\sum\limits_{j=1}^{n_2}\bigg\{\frac{g'(\mu_1)A_{2j}(Y_{2j}-\mu_1)}{\pi_2}-\frac{g'(\mu_0)(1-A_{2j})(Y_{2j}-\mu_0)}{1-\pi_2}\bigg\}
=o_p\left(1/\sqrt{n_2}\right)
\end{aligned}
\end{equation*}
and
\begin{equation*}
\frac{1}{\sqrt{n_2}}\sum\limits_{j=1}^{n_2}(A_{2j}-\pi_2)\left\{\widetilde{b}_2(\bmW_{2j})-b_2(\bmW_{2j})\right\}=o_p(1).
\end{equation*}
Therefore, we have
\begin{equation*}
\begin{aligned}
&\sqrt{n_1}\left\{\widehat\delta^{\text{aug}}(\widetilde b_1,\widetilde b_2,\widetilde\theta)-\delta\right\}\\
&=\frac{1}{\sqrt{n_1}}\sum_{i=1}^{n_1}\left\{\frac{\theta g'(\mu_1)A_{1i}(Y_{1i}-\mu_1)}{\pi_1}-\frac{\theta g'(\mu_0)(1-A_{1i})(Y_{1i}-\mu_0)}{1-\pi_1}-(A_{1i}-\pi_1)b_1(\bmW_{1i})\right\}\\
&\quad+\frac{\sqrt{n_1}}{n_2}\sum_{j=1}^{n_2}\bigg\{\frac{(1-\theta)g'(\mu_1)A_{2j}(Y_{2j}-\mu_1)}{\pi_2}-\frac{(1-\theta)g'(\mu_0)(1-A_{2j})(Y_{2j}-\mu_0)}{1-\pi_2}\\
&\qquad\qquad\qquad-(A_{2j}-\pi_2)b_2(\bmW_{2j})\bigg\}
+o_p(1)+o_p\left\{\sqrt{n_1}(\widehat\mu_1-\mu_1)\right\}+o_p\left\{\sqrt{n_1}(\widehat\mu_0-\mu_0)\right\}\\
&=\sqrt{n_1}Z_3+\sqrt{n_1}Z_4+o_p(1),
\end{aligned}
\end{equation*}
where
\begin{equation*}
\begin{aligned}
Z_3&=\frac{1}{n_1}\sum_{i=1}^{n_1}\left\{\frac{\theta g'(\mu_1)A_{1i}(Y_{1i}-\mu_1)}{\pi_1}-\frac{\theta g'(\mu_0)(1-A_{1i})(Y_{1i}-\mu_0)}{1-\pi_1}-(A_{1i}-\pi_1)b_1(\bmW_{1i})\right\},\\
Z_4&=\frac{1}{n_2}\sum_{j=1}^{n_2}\bigg\{\frac{(1-\theta)g'(\mu_1)A_{2j}(Y_{2j}-\mu_1)}{\pi_2}-\frac{(1-\theta)g'(\mu_0)(1-A_{2j})(Y_{2j}-\mu_0)}{1-\pi_2}\\
&\qquad\qquad\qquad-(A_{2j}-\pi_2)b_2(\bmW_{2j})\bigg\},
\end{aligned}
\end{equation*}
and the second equality is due to $\sqrt{n_1}(\widehat\mu_1-\mu_1)=O_p(1)$ and $\sqrt{n_1}(\widehat\mu_0-\mu_0)=O_P(1)$ by the central limit theorem.
%
Following the proofs of Theorem 1, we can show that
\begin{equation*}
\sqrt{n_1}(Z_3+Z_4)\stackrel{d}{\longrightarrow}N\left(0, \var\{\psi_1^{\text{aug}}(\bmO_1)\}+\lambda\var\{\psi_2^{\text{aug}}(\bmO_2^*)\}\right),
\end{equation*}
where
\begin{equation*}
\begin{aligned}
\psi_1^{\text{aug}}(\bmO_1)&=\frac{\theta g'(\mu_1)A_{1}(Y_{1}-\mu_1)}{\pi_1}-\frac{\theta g'(\mu_0)(1-A_{1})(Y_{1}-\mu_0)}{1-\pi_1}-(A_{1}-\pi_1)b_1(\bmW),\\
\psi_2^{\text{aug}}(\bmO_2^*)&=\frac{(1-\theta)g'(\mu_1)A_{2}^*(Y_{2}^*-\mu_1)}{\pi_2^*}-\frac{(1-\theta)g'(\mu_0)(1-A_{2}^*)(Y_{2}^*-\mu_0)}{1-\pi_2^*}-(A_{2}^*-\pi_2^*)b_2(\bmW),
\end{aligned}
\end{equation*}
and $\bmO_2^*=(\bmW,A_2^*,Y_2^*)$ is similar to $\bmO_2$ with $\pi_2^*$ replacing $\pi_2$ and $\pr\{A_2^*=1|\bmW,Y(1),Y(0)\}=\pr(A_2^*=1)=\pi_2^*$. Applying the Slutsky's theorem yields
\begin{equation*}
\sqrt{n_1}\left\{\widehat\delta^{\text{aug}}(\widetilde b_1,\widetilde b_2,\widetilde\theta)-\delta\right\}\stackrel{d}{\longrightarrow} N\left(0, \sigma_{\textsc{cir}}^2(b_1,b_2,\theta)\right).
\end{equation*}

\subsection*{Proof of Theorem S1}

Expanding the function $g(\cdot)$ at $\mu_1$ and $\mu_0$  yields
\begin{equation*}
\begin{aligned}
&\widehat\delta^{\text{aipw}}(c_1,c_2,\eta)-\delta\\
&=\left\{g(\widehat\mu_1^{\text{ipw}})-g(\mu_1)\right\}-\left\{g(\widehat\mu_0^{\text{ipw}})-g(\mu_0)\right\}
-\frac{1}{n_1}\sum_{i=1}^{n_1}\{A_{1i}-p_1(\bmW_{1i})\}c_1(\bmW_{1i})\\
&\quad-\frac{1}{n_2}\sum_{j=1}^{n_2}\{A_{2j}-p_2(\bmW_{2j})\}c_2(\bmW_{2j})\\
&=\eta g'(\mu_1)\left\{\widehat\mu_{11}^{\text{ipw}}-\mu_1\right\}-\eta g'(\mu_0)\left\{\widehat\mu_{10}^{\text{ipw}}-\mu_0\right\}-\frac{1}{n_1}\sum_{i=1}^{n_1}\{A_{1i}-p_1(\bmW_{1i})\}c_1(\bmW_{1i})\\
&\quad+(1-\eta)g'(\mu_1)\left\{\widehat\mu_{21}^{\text{ipw}}-\mu_1\right\}-(1-\eta) g'(\mu_0)\left\{\widehat\mu_{20}^{\text{ipw}}-\mu_0\right\}\\
&\quad-\frac{1}{n_2}\sum_{j=1}^{n_2}\{A_{2j}-p_2(\bmW_{2j})\}c_2(\bmW_{2j})+O_p\left\{(\widehat\mu_1^{\text{ipw}}-\mu_1)^2\right\}+O_p\left\{(\widehat\mu_0^{\text{ipw}}-\mu_0)^2\right\}\\
&= H_1+H_2+O_p\left\{(\widehat\mu_1^{\text{ipw}}-\mu_1)^2\right\}+O_p\left\{(\widehat\mu_0^{\text{ipw}}-\mu_0)^2\right\},
\end{aligned}
\end{equation*}
where
\begin{equation*}
\begin{aligned}
H_1&=\eta g'(\mu_1)\left\{\widehat\mu_{11}^{\text{ipw}}-\mu_1\right\}-\eta g'(\mu_0)\left\{\widehat\mu_{10}^{\text{ipw}}-\mu_0\right\}-\frac{1}{n_1}\sum_{i=1}^{n_1}\{A_{1i}-p_1(\bmW_{1i})\}c_1(\bmW_{1i}),\\
H_2&=(1-\eta)g'(\mu_1)\left\{\widehat\mu_{21}^{\text{ipw}}-\mu_1\right\}-(1-\eta) g'(\mu_0)\left\{\widehat\mu_{20}^{\text{ipw}}-\mu_0\right\}\\
&\quad-\frac{1}{n_2}\sum_{j=1}^{n_2}\{A_{2j}-p_2(\bmW_{2j})\}c_2(\bmW_{2j}).
\end{aligned}
\end{equation*}
Recall that
{\small
\begin{equation*}
\begin{aligned}
&\widehat\mu_{11}^{\text{ipw}}=\bigg\{\sum\limits_{i=1}^{n_1}\frac{I(A_{1i}=1)}{p_1(\bmW_{1i})}\bigg\}^{-1}\bigg\{\sum\limits_{i=1}^{n_1}\frac{I(A_{1i}=1)Y_{1i}}{p_1(\bmW_{1i})}\bigg\},\\
&\widehat\mu_{10}^{\text{ipw}}=\bigg\{\sum\limits_{i=1}^{n_1}\frac{I(A_{1i}=0)}{1-p_1(\bmW_{1i})}\bigg\}^{-1}\bigg\{\sum\limits_{i=1}^{n_1}\frac{I(A_{1i}=0)Y_{1i}}{1-p_1(\bmW_{1i})}\bigg\},\\
&\widehat\mu_{21}^{\text{ipw}}=\bigg\{\sum\limits_{j=1}^{n_2}\frac{I(A_{2j}=1)}{p_2(\bmW_{2j})}\bigg\}^{-1}\bigg\{\sum\limits_{j=1}^{n_2}\frac{I(A_{2j}=1)Y_{2j}}{p_2(\bmW_{2j})}\bigg\}, \\ &\widehat\mu_{20}^{\text{ipw}}=\bigg\{\sum\limits_{j=1}^{n_2}\frac{I(A_{2j}=0)}{1-p_2(\bmW_{2j})}\bigg\}^{-1}\bigg\{\sum\limits_{j=1}^{n_2}\frac{I(A_{2j}=0)Y_{2j}}{1-p_2(\bmW_{2j})}\bigg\}.
\end{aligned}
\end{equation*}}
By the law of large numbers, as $n_1\rightarrow\infty$, we have
\begin{equation*}
\begin{aligned}
&\frac{1}{n_1}\sum\limits_{i=1}^{n_1}\frac{I(A_{1i}=1)}{p_1(\bmW_{1i})}\rightarrow\epn\left\{\frac{I(A_1=1)}{p_1(\bmW)}\right\}=\epn\left[\epn\left\{\frac{I(A_1=1)}{p_1(\bmW)}\bigg|\bmW\right\}\right]=1,\\
&\frac{1}{n_1}\sum\limits_{i=1}^{n_1}\frac{I(A_{1i}=0)}{1-p_1(\bmW_{1i})}\rightarrow\epn\left\{\frac{I(A_1=0)}{1-p_1(\bmW)}\right\}=\epn\left[\epn\left\{\frac{I(A_1=0)}{1-p_1(\bmW)}\bigg|\bmW\right\}\right]=1,
\end{aligned}
\end{equation*}
so that
\begin{equation*}
\frac{1}{n_1}\sum\limits_{i=1}^{n_1}\frac{I(A_{1i}=1)}{p_1(\bmW_{1i})}=1+o_p(1)\quad\text{and}\quad
\frac{1}{n_1}\sum\limits_{i=1}^{n_1}\frac{I(A_{1i}=0)}{1-p_1(\bmW_{1i})}=1+o_p(1).
\end{equation*}
Then we can write
\begin{equation*}
\begin{aligned}
H_1&=\eta g'(\mu_1)\left\{\sum\limits_{i=1}^{n_1}\frac{I(A_{1i}=1)}{p_1(\bmW_{1i})}\right\}^{-1}\left\{\sum\limits_{i=1}^{n_1}\frac{I(A_{1i}=1)(Y_{1i}-\mu_1)}{p_1(\bmW_{1i})}\right\}\\
&\quad-\eta g'(\mu_0)\left\{\sum\limits_{i=1}^{n_1}\frac{I(A_{1i}=0)}{1-p_1(\bmW_{1i})}\right\}^{-1}\left\{\sum\limits_{i=1}^{n_1}\frac{I(A_{1i}=0)(Y_{1i}-\mu_0)}{1-p_1(\bmW_{1i})}\right\}\\
&\quad-\frac{1}{n_1}\sum_{i=1}^{n_1}\{A_{1i}-p_1(\bmW_{1i})\}c_1(\bmW_{1i}),\\
&=\frac{1}{n_1}\sum\limits_{i=1}^{n_1}\frac{\eta g'(\mu_1)A_{1i}(Y_{1i}-\mu_1)}{p_1(\bmW_{1i})}\{1+o_p(1)\}-\frac{1}{n_1}\sum\limits_{i=1}^{n_1}\frac{\eta g'(\mu_0)(1-A_{1i})(Y_{1i}-\mu_0)}{1-p_1(\bmW_{1i})}\{1+o_p(1)\}\\
&\quad-\frac{1}{n_1}\sum_{i=1}^{n_1}\{A_{1i}-p_1(\bmW_{1i})\}c_1(\bmW_{1i}),\\
&=\frac{1}{n_1}\sum\limits_{i=1}^{n_1}\left[\frac{\eta g'(\mu_1)A_{1i}(Y_{1i}-\mu_1)}{p_1(\bmW_{1i})}-\frac{\eta g'(\mu_0)(1-A_{1i})(Y_{1i}-\mu_0)}{1-p_1(\bmW_{1i})}-\{A_{1i}-p_1(\bmW_{1i})\}c_1(\bmW_{1i})\right]\\
&\qquad\qquad\times\{1+o_p(1)\}\\
&=\frac{1}{n_1}\sum\limits_{i=1}^{n_1}\psi_1^{\text{aipw}}(\bmO_{1i})\{1+o_p(1)\},
\end{aligned}
\end{equation*}
where
\begin{equation*}
\psi_1^{\text{aipw}}(\bmO_{1})=\frac{\eta g'(\mu_1)A_{1}(Y_{1}-\mu_1)}{p_1(\bmW)}-\frac{\eta g'(\mu_0)(1-A_{1})(Y_{1}-\mu_0)}{1-p_1(\bmW)}-\{A_{1}-p_1(\bmW)\}c_1(\bmW).
\end{equation*}
The central limit theorem ensures that, as $n_1\rightarrow\infty$,
\begin{equation}\label{stage1AsyDis.cdr}\tag{A.3}
\sqrt{n_1}H_1\stackrel{d}{\longrightarrow} N\left(0, \var\{\psi_1^{\text{aipw}}(\bmO_1)\}\right).
\end{equation}
By conditioning on the stage 1 data $\bmD_1$,  we have
\begin{equation*}
\frac{1}{n_2}\sum\limits_{j=1}^{n_2}\frac{I(A_{2j}=1)}{p_2(\bmW_{2j})}=1+o_p(1)\quad\text{and}\quad
\frac{1}{n_2}\sum\limits_{j=1}^{n_2}\frac{I(A_{2j}=0)}{1-p_2(\bmW_{2j})}=1+o_p(1),
\end{equation*}
as $n_2\rightarrow\infty$.
Then $H_2$ can be written as
\begin{equation*}
\begin{aligned}
H_2&=\frac{1}{n_2}\sum\limits_{j=1}^{n_2}\frac{(1-\eta)g'(\mu_1)A_{2j}(Y_{2j}-\mu_1)}{p_2(\bmW_{2j})}\{1+o_p(1)\}\\
&\quad-\frac{1}{n_2}\sum\limits_{j=1}^{n_2}\frac{(1-\eta)g'(\mu_0)(1-A_{2j})(Y_{2j}-\mu_0)}{1-p_2(\bmW_{2j})}\{1+o_p(1)\}-\frac{1}{n_2}\sum_{j=1}^{n_2}\{A_{2j}-p_2(\bmW_{2j})\}c_2(\bmW_{2j})\\
&=\frac{1}{n_2}\sum\limits_{j=1}^{n_2}\bigg[\frac{(1-\eta)g'(\mu_1)A_{2j}(Y_{2j}-\mu_1)}{p_2(\bmW_{2j})}-\frac{(1-\eta)g'(\mu_0)(1-A_{2j})(Y_{2j}-\mu_0)}{1-p_2(\bmW_{2j})} \\
&\qquad\qquad\quad-\{A_{2j}-p_2(\bmW_{2j})\}c_2(\bmW_{2j})\bigg]\{1+o_p(1)\}\\
&=\frac{1}{n_2}\sum\limits_{j=1}^{n_2}\psi_2^{\text{aipw}}(\bmO_{2j})\{1+o_p(1)\},
\end{aligned}
\end{equation*}
where
\begin{equation*}
\psi_2^{\text{aipw}}(\bmO_{2})=\frac{(1-\eta)g'(\mu_1)A_{2}(Y_{2}-\mu_1)}{p_2(\bmW)}-\frac{(1-\eta)g'(\mu_0)(1-A_{2})(Y_{2}-\mu_0)}{1-p_2(\bmW)}-\{A_{2}-p_2(\bmW)\}c_2(\bmW).
\end{equation*}
By the central limit theorem, we have
\begin{equation}\label{stage2AsyDis.cdr}\tag{A.4}
\sqrt{n_2}H_2 \mid \bmD_1\stackrel{d}{\longrightarrow} N\left(0, \var\{\psi_2^{\text{aipw}}(\bmO_2)\}\right),\quad\text{as }n_2\rightarrow\infty.
\end{equation}

Moreover, we observe that
\begin{equation*}
\begin{aligned}
&\pr(\sqrt{n_1}H_1\leq h_1, \sqrt{n_2}H_2\leq h_2)\\
&=\pr(\sqrt{n_1}H_1\leq h_1)\pr(\sqrt{n_2}H_2\leq h_2\,|\,\sqrt{n_1}H_1\leq h_1)\\
&=\pr(\sqrt{n_1}H_1\leq h_1)\epn\left\{I(\sqrt{n_2}H_2\leq h_2)\;|\;\sqrt{n_1}H_1\leq h_1\right\}\\
&=\pr(\sqrt{n_1}H_1\leq h_1)\epn\left[\epn\left\{I(\sqrt{n_2}H_2\leq h_2) | \bmD_1, \sqrt{n_1}H_1\leq h_1\right\} | \sqrt{n_1}H_1\leq h_1\right]\\
&=\pr(\sqrt{n_1}H_1\leq h_1)\epn\left\{\pr(\sqrt{n_2}H_2\leq h_2 | \bmD_1) | \sqrt{n_1}H_1\leq h_1\right\}
\end{aligned}
\end{equation*}
By \eqref{stage2AsyDis.cdr}, as $n_2\rightarrow\infty$, $\pr(\sqrt{n_2}H_2\leq h_2|\bmD_1)\rightarrow\Phi(h_2/\sqrt{\var\{\psi_2^{\text{aipw}}(\bmO_2)\}})$. Further letting $n_1\rightarrow\infty$, we have $\pr(\sqrt{n_1}H_1\leq h_1)\rightarrow\Phi(h_1/\sqrt{\var\{\psi_1^{\text{aipw}}(\bmO_1)\}})$ and $\Phi(h_2/\sqrt{\var\{\psi_2^{\text{aipw}}(\bmO_2)\}})\rightarrow\Phi(h_2/\sqrt{\var\{\psi_2^{\text{aipw}}(\bmO_2^*)\}})$, where
\begin{equation*}
\begin{aligned}
\psi_2^{\text{aipw}}(\bmO_2^*)&=\frac{(1-\eta)g'(\mu_1)A_{2}^*(Y_{2}^*-\mu_1)}{p_2^*(\bmW)}-\frac{(1-\eta)g'(\mu_0)(1-A_{2}^*)(Y_{2}^*-\mu_0)}{1-p_2^*(\bmW)}\\
&\quad-\{A_{2}^*-p_2^*(\bmW)\}c_2(\bmW)
\end{aligned}
\end{equation*}
and $\bmO_2^*=(\bmW,A_2^*,Y_2^*)$ is similar to $\bmO_2$ with $p_2^*$ replacing $p_2$,
$\pr\{A_2^*=1|\bmW,Y(1),Y(0)\}=\pr(A_2^*=1|\bmW)=p_2^*(\bmW)$, and
$Y_2^*=Y(A_2^*)=A_2^*Y(1)+(1-A_2^*)Y(0)$. This, together with the condition that $n_1/n_2\rightarrow\lambda$ as $n_1\rightarrow\infty$, yields
\begin{equation*}
\pr(\sqrt{n_1}H_1\leq h_1, \sqrt{n_2}H_2\leq h_2)\rightarrow\Phi\left(\frac{h_1}{\sqrt{\var\{\psi_1^{\text{aipw}}(\bmO_1)\}}}\right)\Phi\left(\frac{h_2}{\sqrt{\var\{\psi_2^{\text{aipw}}(\bmO_2^*)\}}}\right).
\end{equation*}
It thus follows that
\begin{equation*}
\sqrt{n_1}\begin{pmatrix}H_1\\H_2\end{pmatrix}\stackrel{d}{\longrightarrow}N\left(\begin{pmatrix}0\\0\end{pmatrix}, \begin{pmatrix}\var\{\psi_1^{\text{aipw}}(\bmO_1)\}&0\\0&\lambda\var\{\psi_2^{\text{aipw}}(\bmO_2; p_2^*)\}\end{pmatrix}\right).
\end{equation*}
The delta method ensures that
$$\sqrt{n_1}(H_1+H_2)\stackrel{d}{\longrightarrow}N\left(0, \var\{\psi_1^{\text{aipw}}(\bmO_1)\}+\lambda\var\{\psi_2^{\text{aipw}}(\bmO_2^*)\}\right)$$
and hence
\begin{equation*}
\sqrt{n_1}\left\{\widehat\delta^{\text{aipw}}(c_1,c_2,\eta)-\delta\right\}\stackrel{d}{\longrightarrow} N\left(0, \var\{\psi_1^{\text{aipw}}(\bmO_1)\}+\lambda\var\{\psi_2^{\text{aipw}}(\bmO_2^*)\}\right).
\end{equation*}

\subsection*{Derivation of  $(c_{1,\text{opt}},c_{2,\text{opt}})$}

The asymptotic variance of the AIPW estimator $\widehat\delta^{\text{aipw}}(c_1,c_2,\eta)$ is $$\sigma_{\textsc{cdr}}^2(c_1,c_2,\eta)=\var\{\psi_1^{\text{aipw}}(\bmO_1)\}+\lambda\var\{\psi_2^{\text{aipw}}(\bmO_2^*)\}.$$
We observe that $\var\{\psi_1^{\text{aipw}}(\bmO_1)\}$ (rsp. $\var\{\psi_2^{\text{aipw}}(\bmO_2^*)\}$) only depends on $c_1$ (rsp. $c_2$).  To minimize $\var\{\psi_1^{\text{aipw}}(\bmO_1)\}+\lambda\var\{\psi_2^{\text{aipw}}(\bmO_2^*)\}$ with respect to $c_1$ (rsp. $c_2$), it suffices to consider $\var\{\psi_1^{\text{aipw}}(\bmO_1)\}$ (rsp. $\var\{\psi_2^{\text{aipw}}(\bmO_2; p_2^*)\}$). By conditioning on $\bmW$, we can write
\begin{equation*}
\begin{aligned}
\var\{\psi_1^{\text{aipw}}(\bmO_1)\}&=\epn\left(\epn\left[\left\{\frac{\eta g'(\mu_1)A_{1}(Y_{1}-\mu_1)}{p_1(\bmW)}-\frac{\eta g'(\mu_0)(1-A_{1})(Y_{1}-\mu_0)}{1-p_1(\bmW)}\right\}^2\bigg| \bmW\right]\right)\\
&\quad-2\epn\left(\epn\left[\eta\{A_1-p_1(\bmW)\}g'(\mu_1)A_1(Y_1-\mu_1) \bigg|\bmW\right]\frac{c_1(\bmW)}{p_1(\bmW)}\right)\\
&\quad+2\epn\left(\epn\left[\eta\{A_1-p_1(\bmW)\}g'(\mu_0)(1-A_1)(Y_1-\mu_0) \bigg|\bmW\right]\frac{c_1(\bmW)}{1-p_1(\bmW)}\right)\\
&\quad+\epn\left(\epn\left[\{A_1-p_1(\bmW)\}^2|\bmW\right]c_1(\bmW)^2\right).
\end{aligned}
\end{equation*}
The minimizer of $\var\{\psi_1^{\text{aipw}}(\bmO_1)\}$ with respect to $c_1$ is found to be
\begin{equation*}
\begin{aligned}
c_{1,\text{opt}}(\bmW;\eta)&=\frac{2\epn\left[\eta\{A_1-p_1(\bmW)\}g'(\mu_1)A_1(Y_1-\mu_1)/p_1(\bmW) |\bmW\right]}{2\epn\left[\{A_1-p_1(\bmW)\}^2|\bmW\right]}\\
&\quad-\frac{2\epn\left[\eta\{A_1-p_1(\bmW)\}g'(\mu_0)(1-A_1)(Y_1-\mu_0)/\{1-p_1(\bmW)\} |\bmW\right]}{2\epn\left[\{A_1-p_1(\bmW)\}^2|\bmW\right]}\\
&=\eta\left[\frac{g'(\mu_1)\{m_1(\bmW)-\mu_1\}}{p_1(\bmW)}+
\frac{g'(\mu_0)\{m_0(\bmW)-\mu_0\}}{1-p_1(\bmW)}\right].
\end{aligned}
\end{equation*}
Moreover, by conditioning on $\bmW$, the variance of $\psi_2^{\text{aipw}}(\bmO_2^*)$  is equal to
\begin{equation*}
\begin{aligned}
\var\{\psi_2^{\text{aipw}}(\bmO_2^*)\}&=\epn\left(\epn\left[\left\{\frac{(1-\eta) g'(\mu_1)A_{2}(Y_{2}-\mu_1)}{p_2^*(\bmW)}-\frac{(1-\eta) g'(\mu_0)(1-A_{2})(Y_{2}-\mu_0)}{1-p_2^*(\bmW)}\right\}^2\bigg| \bmW\right]\right)\\
&\quad-2\epn\left(\epn\left[(1-\eta)\{A_2-p_2^*(\bmW)\}g'(\mu_1)A_2(Y_2-\mu_1) \bigg|\bmW\right]\frac{c_2(\bmW)}{p_2^*(\bmW)}\right)\\
&\quad+2\epn\left(\epn\left[(1-\eta)\{A_2-p_2^*(\bmW)\}g'(\mu_0)(1-A_2)(Y_2-\mu_0) \bigg|\bmW\right]\frac{c_2(\bmW)}{1-p_2^*(\bmW)}\right)\\
&\quad+\epn\left(\epn\left[\{A_2-p_2^*(\bmW)\}^2|\bmW\right]c_2(\bmW)^2\right).
\end{aligned}
\end{equation*}
The minimizer of the preceding display with respect to $c_2$ is given by
\begin{equation*}
\begin{aligned}
c_{2,\text{opt}}(\bmW;\eta)&=\frac{2\epn\left[(1-\eta)\{A_2-p_2^*(\bmW)\}g'(\mu_1)A_2(Y_2-\mu_1)/p_2^*(\bmW) |\bmW\right]}{2\epn\left[\{A_2-p_2^*(\bmW)\}^2|\bmW\right]}\\
&\quad-\frac{2\epn\left[(1-\eta)\{A_2-p_2^*(\bmW)\}g'(\mu_0)(1-A_2)(Y_2-\mu_0)/\{1-p_2^*(\bmW)\} |\bmW\right]}{2\epn\left[\{A_2-p_2^*(\bmW)\}^2|\bmW\right]}\\
&=(1-\eta)\left[\frac{g'(\mu_1)\{m_1(\bmW)-\mu_1\}}{p_2^*(\bmW)}+
\frac{g'(\mu_0)\{m_0(\bmW)-\mu_0\}}{1-p_2^*(\bmW)}\right].
\end{aligned}
\end{equation*}

\subsection*{Derivation of $\eta_{\text{opt}}^*$}

Substituting $(c_1^*,c_2^*)$ into $\psi_1^{\text{aipw}}(\bmO_1)$ and $\psi_2^{\text{aipw}}(\bmO_2^*)$ yields
\begin{equation*}
\begin{aligned}
\psi_1^{\text{aipw}}(\bmO_1; c_1^*)&=\frac{\eta g'(\mu_1)A_1\{Y_1-m_1^*(\bmW_1)\}}{p_1(\bmW_1)}-\frac{\eta g'(\mu_0)(1-A_1)\{Y_1-m_0^*(\bmW_1)\}}{1-p_1(\bmW_1)}\\
&\quad+\eta g'(\mu_1)\left\{m_1^*(\bmW_1)-\mu_1\right\}-\eta g'(\mu_0)\left\{m_0^*(\bmW_1)-\mu_0\right\},\\
\psi_2^{\text{aipw}}(\bmO_2^*,c_2^*)&=\frac{(1-\eta) g'(\mu_1)A_2\{Y_2-m_1^*(\bmW_2)\}}{p_2^*(\bmW_2)}-\frac{(1-\eta) g'(\mu_0)(1-A_2)\{Y_2-m_0^*(\bmW_2)\}}{1-p_2^*(\bmW_2)}\\
&\quad+(1-\eta)g'(\mu_1)\left\{m_1^*(\bmW_2)-\mu_1\right\}-(1-\eta) g'(\mu_0)\left\{m_0^*(\bmW_2)-\mu_0\right\}.
\end{aligned}
\end{equation*}
The  asymptotic variance $\sigma_{\textsc{cdr}}^2(c_1^*,c_2^*,\eta)$ can then be written as
\begin{equation*}
\begin{aligned}
\sigma_{\textsc{cdr}}^2(c_1^*,c_2^*,\eta)&=
\var\{\psi_1^{\text{aipw}}(\bmO_1; c_1^*)\}+\lambda\var\{\psi_2^{\text{aipw}}(\bmO_2^*; c_2^*)\}\\
&=\eta^2\sigma_{1,\textsc{cdr}}^{*2}+\lambda(1-\eta)^2\sigma_{2,\textsc{cdr}}^{*2},
\end{aligned}
\end{equation*}
where
\begin{equation*}
\begin{aligned}
\sigma_{1,\textsc{cdr}}^{*2}&=\var\bigg[\frac{g'(\mu_1)A_1\{Y_1-m_1^*(\bmW)\}}{p_1(\bmW)}-\frac{g'(\mu_0)(1-A_1)\{Y_1-m_0^*(\bmW)\}}{1-p_1(\bmW)}\\
&\qquad\quad+g'(\mu_1)\left\{m_1^*(\bmW)-\mu_1\right\}-g'(\mu_0)\left\{m_0^*(\bmW)-\mu_0\right\}\bigg],\\
\sigma_{2,\textsc{cdr}}^{*2}&=\var\bigg[\frac{g'(\mu_1)A_2^*\{Y_2^*-m_1^*(\bmW)\}}{p_2^*(\bmW)}-\frac{g'(\mu_0)(1-A_2^*)\{Y_2^*-m_0^*(\bmW)\}}{1-p_2^*(\bmW)}\\
&\qquad\quad+g'(\mu_1)\left\{m_1^*(\bmW)-\mu_1\right\}-g'(\mu_0)\left\{m_0^*(\bmW)-\mu_0\right\}\bigg].\\
\end{aligned}
\end{equation*}
The minimizer of $\var\{\psi_1^{\text{aipw}}(\bmO_1; c_1^*)\}+\lambda\var\{\psi_2^{\text{aipw}}(\bmO_2^*; c_2^*)\}$ with respect to $\eta$ is given by
$$\eta_{\text{opt}}^*=\frac{\lambda\sigma_{2,\textsc{cdr}}^{*2}}{\sigma_{1,\textsc{cdr}}^{*2}+\lambda\sigma_{2,\textsc{cdr}}^{*2}}.$$

\subsection*{Proof of Theorem S2}

According to the proofs of Theorem S1, we can write
\begin{equation*}
\begin{aligned}
&\widehat\delta^{\text{aipw}}(\widetilde c_1,\widetilde c_2,\widetilde\eta)-\delta\\
&=\frac{1}{n_1}\sum\limits_{i=1}^{n_1}\bigg[\frac{\eta g'(\mu_1)A_{1i}(Y_{1i}-\mu_1)}{p_1(\bmW_{1i})}+\frac{\eta g'(\mu_0)(1-A_{1i})(Y_{1i}-\mu_0)}{1-p_1(\bmW_{1i})}-\{A_{1i}-p_1(\bmW_{1i})\}c_1(\bmW_{1i})\bigg]\\
&\quad+\frac{1}{n_2}\sum\limits_{j=1}^{n_2}\bigg[\frac{(1-\eta)g'(\mu_1)A_{2j}(Y_{2j}-\mu_1)}{p_2(\bmW_{2j})}-\frac{(1-\eta)g'(\mu_0)(1-A_{2j})(Y_{2j}-\mu_0)}{1-p_2(\bmW_{2j})} \\
&\qquad\qquad\quad-\{A_{2j}-p_2(\bmW_{2j})\}c_2(\bmW_{2j})\bigg]\\
&\quad+\frac{1}{n_1}\sum\limits_{i=1}^{n_1}\bigg[\frac{(\widetilde{\eta}-\eta) g'(\mu_1)A_{1i}(Y_{1i}-\mu_1)}{p_1(\bmW_{1i})}+\frac{(\widetilde{\eta}-\eta) g'(\mu_0)(1-A_{1i})(Y_{1i}-\mu_0)}{1-p_1(\bmW_{1i})}\bigg]\\
&\quad-\frac{1}{n_1}\sum\limits_{i=1}^{n_1}\{A_{1i}-p_1(\bmW_{1i})\}\{\widetilde{c}_1(\bmW_{1i})-c_1(\bmW_{1i})\}\\
&\quad+\frac{1}{n_2}\sum\limits_{j=1}^{n_2}\bigg[\frac{(\eta-\widetilde{\eta})g'(\mu_1)A_{2j}(Y_{2j}-\mu_1)}{p_2(\bmW_{2j})}-\frac{(\eta-\widetilde{\eta})g'(\mu_0)(1-A_{2j})(Y_{2j}-\mu_0)}{1-p_2(\bmW_{2j})}\bigg] \\
&\quad-\frac{1}{n_2}\sum\limits_{j=1}^{n_2}\{A_{2j}-p_2(\bmW_{2j})\}\{\widetilde{c}_2(\bmW_{2j})-c_2(\bmW_{2j})\}+o_p(\widehat\mu_1^{\text{ipw}}-\mu_1)+o_p(\widehat\mu_0^{\text{ipw}}-\mu_0).\\
\end{aligned}
\end{equation*}
The  condition  $\widetilde{\eta}\rightarrow\eta$ as $n_1\rightarrow\infty$ ensures that
\begin{equation*}
\begin{aligned}
&\frac{1}{n_1}\sum\limits_{i=1}^{n_1}\bigg[\frac{(\widetilde{\eta}-\eta) g'(\mu_1)A_{1i}(Y_{1i}-\mu_1)}{p_1(\bmW_{1i})}+\frac{(\widetilde{\eta}-\eta) g'(\mu_0)(1-A_{1i})(Y_{1i}-\mu_0)}{1-p_1(\bmW_{1i})}\bigg]=o_p\left(1/\sqrt{n_1}\right).\\
\end{aligned}
\end{equation*}
By the chebyshev's inequality, for any real number $\varepsilon$, we have
\begin{equation*}
\begin{aligned}
&\pr\left[\frac{1}{\sqrt{n_1}}\sum\limits_{i=1}^{n_1}\{A_{1i}-p_1(\bmW_{1i})\}\{\widetilde{c}_1(\bmW_{1i})-c_1(\bmW_{1i})\}>\varepsilon\right]\\
&\leq\frac{\epn\left[\sum\limits_{i=1}^{n_1}\{A_{1i}-p_1(\bmW_{1i})\}\{\widetilde{c}_1(\bmW_{1i})-c_1(\bmW_{1i})\}\right]^2}{n_1\varepsilon^2}\\
&=\frac{\epn\left[(A_{1i}-p_1(\bmW_{1i}))^2\{\widetilde{c}_1(\bmW_{1i})-c_1(\bmW_{1i})\}^2\right]}{\varepsilon^2}\\
&\leq\frac{\epn\left\{\widetilde{c}_1(\bmW_{1i})-c_1(\bmW_{1i})\right\}^2}{\varepsilon^2}\\
&=o_p(1),
\end{aligned}
\end{equation*}
where the second inequality is because $|A_{1i}-p_1(\bmW_{1i})|\leq1$ and the last equality is because the condition $\|\widetilde{c}_1-c_1\|_2=[\epn\{\widetilde{c}_1-c_1\}^2]^{1/2}=o_p(1)$, Hence
$$\frac{1}{\sqrt{n_1}}\sum\limits_{i=1}^{n_1}\{A_{1i}-p_1(\bmW_{1i})\}\{\widetilde{c}_1(\bmW_{1i})-c_1(\bmW_{1i})\}=o_p(1).$$
By conditioning on the stage 1 data $\bmD_1$, we can similarly obtain that
\begin{equation*}
\begin{aligned}
&\frac{1}{n_2}\sum\limits_{j=1}^{n_2}\bigg[\frac{(\eta-\widetilde{\eta})g'(\mu_1)A_{2j}(Y_{2j}-\mu_1)}{p_2(\bmW_{2j})}-\frac{(\eta-\widetilde{\eta})g'(\mu_0)(1-A_{2j})(Y_{2j}-\mu_0)}{1-p_2(\bmW_{2j})}\bigg]=
o_p\left(1/\sqrt{n_2}\right),\\
&\frac{1}{\sqrt{n_2}}\sum\limits_{j=1}^{n_2}\{A_{2j}-p_2(\bmW_{2j})\}\{\widetilde{c}_2(\bmW_{2j})-c_2(\bmW_{2j})\}=o_p(1).
\end{aligned}
\end{equation*}
Combining these results leads to
\begin{equation*}
\begin{aligned}
&\sqrt{n_1}\left\{\widehat\delta^{\text{aipw}}(\widetilde{c}_1,\widetilde{c}_2,\widetilde{\eta})-\delta\right\}\\
&=\frac{1}{\sqrt{n_1}}\sum\limits_{i=1}^{n_1}\bigg[\frac{\eta g'(\mu_1)A_{1i}(Y_{1i}-\mu_1)}{p_1(\bmW_{1i})}+\frac{\eta g'(\mu_0)(1-A_{1i})(Y_{1i}-\mu_0)}{1-p_1(\bmW_{1i})}-\{A_{1i}-p_1(\bmW_{1i})\}c_1(\bmW_{1i})\bigg]\\
&\quad+\frac{\sqrt{n_1}}{n_2}\sum\limits_{j=1}^{n_2}\bigg[\frac{(1-\eta)g'(\mu_1)A_{2j}(Y_{2j}-\mu_1)}{p_2(\bmW_{2j})}-\frac{(1-\eta)g'(\mu_0)(1-A_{2j})(Y_{2j}-\mu_0)}{1-p_2(\bmW_{2j})} \\
&\qquad\qquad\quad-\{A_{2j}-p_2(\bmW_{2j})\}c_2(\bmW_{2j})\bigg]\\
&\quad+o_p(1)+o_p\{\sqrt{n_1}(\widehat\mu_1^{\text{ipw}}-\mu_1)\}+o_p\{\sqrt{n_1}(\widehat\mu_0^{\text{ipw}}-\mu_0)\}\\
&= \sqrt{n_1}H_3+\sqrt{n_1}H_4+o_p(1),
\end{aligned}
\end{equation*}
where
\begin{equation*}
\begin{aligned}
H_3&=\frac{1}{n_1}\sum\limits_{i=1}^{n_1}\bigg[\frac{\eta g'(\mu_1)A_{1i}(Y_{1i}-\mu_1)}{p_1(\bmW_{1i})}+\frac{\eta g'(\mu_0)(1-A_{1i})(Y_{1i}-\mu_0)}{1-p_1(\bmW_{1i})}-\{A_{1i}-p_1(\bmW_{1i})\}c_1(\bmW_{1i})\bigg],\\
H_4&=\frac{1}{n_2}\sum\limits_{j=1}^{n_2}\bigg[\frac{(1-\eta)g'(\mu_1)A_{2j}(Y_{2j}-\mu_1)}{p_2(\bmW_{2j})}-\frac{(1-\eta)g'(\mu_0)(1-A_{2j})(Y_{2j}-\mu_0)}{1-p_2(\bmW_{2j})}\\
&\qquad\qquad\quad-\{A_{2j}-p_2(\bmW_{2j})\}c_2(\bmW_{2j})\bigg].
\end{aligned}
\end{equation*}
Following the proofs of Theorem S1, we can show that
\begin{equation*}
\sqrt{n_1}(H_3+H_4)\stackrel{d}{\longrightarrow} N\left(0, \var\{\psi_1^{\text{aipw}}(\bmO_1)\}+\lambda\var\{\psi_2^{\text{aipw}}(\bmO_2^*)\}\right),
\end{equation*}
where
\begin{equation*}
\begin{aligned}
\psi_1^{\text{aipw}}(\bmO_1)&=\frac{\eta g'(\mu_1)A_{1}(Y_{1}-\mu_1)}{p_1(\bmW)}-\frac{\eta g'(\mu_0)(1-A_{1})(Y_{1}-\mu_0)}{1-p_1(\bmW)}-\{A_{1}-p_1(\bmW)\}c_1(\bmW),\\
\psi_2^{\text{aipw}}(\bmO_2^*)&=\frac{(1-\eta)g'(\mu_1)A_{2}^*(Y_{2}^*-\mu_1)}{p_2^*(\bmW)}-\frac{(1-\eta)g'(\mu_0)(1-A_{2}^*)(Y_{2}^*-\mu_0)}{1-p_2^*(\bmW)}\\
&\quad-\{A_{2}^*-p_2^*(\bmW)\}c_2(\bmW),
\end{aligned}
\end{equation*}
and $\bmO_2^*=(\bmW,A_2^*,Y_2^*)$ is similar to $\bmO_2$ with $p_2^*$ replacing $p_2$,
$\pr\{A_2^*=1|\bmW,Y(1),Y(0)\}=\pr(A_2^*=1|\bmW)=p_2^*(\bmW)$, and $Y_2^*=Y(A_2^*)=A_2^*Y(1)+(1-A_2^*)Y(0)$.
The conclusion of the theorem thus follows.

\section*{Appendix D: Additional Simulations}\label{add.simu}

\subsection*{Smaller sample size}\label{small.samsize}

Table \ref{sim.sss.re} reports additional simulation results for a smaller sample size ($n_1=n_0=150$). Except for the reduced sample size, this simulation study is identical to the one described in Section 5. We will focus on optimized estimators in design comparisons. In Setting 1, the two-stage designs perform similarly to each other and generally better than the one-stage design. This suggests that stage 1 data with $n_1 = 150$ may be insufficient for estimating the optimal CDR design. In Setting 2, the results for $n_1=n_0=150$ are similar to those for $n_1=n_0=250$. Specifically, the one-stage CDR design continues to underperform  the one-stage CIR design, and the two-stage designs consistently outperform the one-stage designs. Among the three two-stage designs, the two-stage hybrid design generally achieves the highest efficiency, although the incremental improvement tends to be small.

\subsection*{Allocation of stage-specific sample sizes}\label{vary.samsize}

Here we report a simulation study to evaluate how the sample size allocation between stages affects the performance of adaptive designs. Fixing the total sample size at $n=500$, we  consider two alternative proportions for stage 1 sample size:  $n_1/n=30\%$ and $n_1/n=70\%$, in addition to the original scenario with $n_1/n=50\%$. In all other aspects, this simulation study is identical to Setting 1 of the main simulation study (described in Section 5.1).

Table \ref{sim.vary.samsize} reports relative efficiency results with the optimized estimator under the one-stage design as the reference. When $n_1/n=30\%$, the two-stage CIR and CDR designs outperform the one-stage design, with the two-stage CIR design achieving the highest efficiency. This suggests that a stage 1 sample size of $n_1=150$ may be insufficient to fully optimize the CDR design. When $n_1/n=70\%$, the two-stage CIR and CDR designs perform similarly and both remain more efficient than the one-stage design. Although a larger stage 1 sample size ($n_1=350$) improves estimation of the optimal CDR design, the reduced Stage 2 sample limits the impact of the estimated optimal design. These results highlight the trade-off between allocating sufficient stage 1 data for optimizing treatment allocation and retaining adequate Stage 2 data for precise treatment effect estimation.

\subsection*{Non-prognostic covariates}\label{rob.nonprogcov}

Here, we present a simulation study in situations where some baseline covariates are not prognostic. Following Setting 1 in Section 5.1 of the main text, we generate the potential outcomes using the logistic regression model
$\logit[\pr\{Y(a)=1|\bmW\}]=\gamma_0+\gamma_1a+\bmgamma_2'\bmW+\bmgamma_3'(a\bmW),$
where $\gamma_0=-2.5$ and $\gamma_1=1$ or $2$, for $a=0,1$. We consider two scenarios that differ in the number of prognostic covariates: (1) one prognostic covariate, with $\bmgamma_2=(0,0,0.2)'$ and $\bmgamma_3=(0,0,-1.5)'$; and (2) two prognostic covariates, with $\bmgamma_2=(0,-0.2,0.2)'$ and $\bmgamma_3=(0,-1,-1.5)'$. In all other aspects, this simulation study is identical to Setting 1 of the main study.

Table \ref{sim.nonprogcov} reports the relative efficiency results using the optimized estimator under the one-stage CIR design as the reference. The results resemble those shown in the upper section of Table 1 in the main text. The two-stage CIR and CDR designs are generally more efficient than the one-stage design. The efficiency gains from the two-stage designs tend to increase with the number of prognostic covariates, though the improvements are small.  When $\bmX$ includes at least one prognostic covariate, the two-stage CDR design typically achieves the highest efficiency among the three designs. In cases where $\bmX$ is not prognostic for either treatment, for example, in Scenario (1) with $\bmX=W_1, W_2$ or $(W_1,W_2)'$, the two-stage CIR and CDR designs perform similarly. In these cases, the optimal treatment allocation reduces to the Neyman allocation for both CIR and CDR, which explains why the two-stage designs perform similarly to each other and better than the one-stage design.

\subsection*{More severe model misspecification}\label{severe.misspec}

The model misspecification in the main simulation study is somewhat moderate. To investigate the impact of more severe misspecification, we have conducted an additional simulation study, which is identical to Setting 1 of the main study except for a different working model for design optimization: $\logit\{\pr(Y_1=1|A_1,\bmX)\}=\alpha_0+\alpha_1A_1+\bmalpha_2'\bmX$. This working model is obviously more severely misspecified than the original model due to the omission of the treatment-by-covariate interaction terms. The results of this additional study, shown in Table S5, are generally consistent with the original results in that the two-stage CIR and CDR designs continue to outperform the one-stage design. A notable difference is that the efficiency advantage of the two-stage CDR design over the two-stage CIR design appears to have diminished, particularly when the dimension of $\bmX$ is larger (2 or 3). The two-stage CIR design appears more robust against this more severe form of misspecification.

\pagebreak
\begin{landscape}
\renewcommand{\baselinestretch}{1.1}
\begin{table}[htbp]
{\footnotesize
\caption{Additional simulation results: (standard deviation; median standard error) for optimized estimators under various two-stage (2S) designs.}\label{sim.rst.sdse}
\newcolumntype{d}{D{.}{.}{2}}
\newcolumntype{e}{D{.}{.}{1}}
\begin{center}
\begin{tabular}{cccccccccc}
\hline
\hline
Setting&$\gamma_1$&Design&\multicolumn{7}{c}{$\bmX$}\\
\cline{4-10}
&&&\multicolumn{1}{c}{$W_1$}&\multicolumn{1}{c}{$W_2$}&\multicolumn{1}{c}{$W_3$}&\multicolumn{1}{c}{$(W_1,W_2)'$}&\multicolumn{1}{c}{$(W_1,W_3)'$}&\multicolumn{1}{c}{$(W_2,W_3)'$}&\multicolumn{1}{c}{$\bmW$}\\
\hline
1&1&2S CIR    &(0.253, 0.253)&(0.252, 0.253)&(0.252, 0.252)&(0.252, 0.252)&(0.252, 0.252)&(0.251, 0.252)&(0.250, 0.252)\\
  &&2S CDR    &(0.248, 0.253)&(0.250, 0.253)&(0.246, 0.252)&(0.248, 0.252)&(0.246, 0.251)&(0.246, 0.251)&(0.246, 0.250)\\
\cline{2-10}
 &2&2S CIR    &(0.249, 0.248)&(0.248, 0.248)&(0.248, 0.248)&(0.248, 0.248)&(0.248, 0.248)&(0.247, 0.248)&(0.247,0.248)\\
  &&2S CDR    &(0.245, 0.248)&(0.246, 0.248)&(0.245, 0.248)&(0.244, 0.248)&(0.245, 0.247)&(0.244, 0.247)&(0.243, 0.246)\\
\hline
2&1&2S CIR    &(0.246, 0.245)&(0.243, 0.245)&(0.246, 0.244)&(0.247, 0.244)&(0.249, 0.244)&(0.245, 0.244)&(0.245, 0.243)\\
  &&2S CDR    &(0.251, 0.246)&(0.248, 0.246)&(0.247, 0.244)&(0.248, 0.246)&(0.249, 0.243)&(0.251, 0.243)&(0.255, 0.242)\\
  &&2S Hybrid &(0.245, 0.245)&(0.243, 0.245)&(0.242, 0.244)&(0.247, 0.244)&(0.243, 0.242)&(0.244, 0.242)&(0.240, 0.241)\\
\cline{2-10}
 &2&2S CIR    &(0.242, 0.240)&(0.239, 0.239)&(0.241, 0.239)&(0.239, 0.238)&(0.240, 0.239)&(0.237, 0.238)&(0.242, 0.238)\\
  &&2S CDR    &(0.245, 0.241)&(0.241, 0.240)&(0.242, 0.239)&(0.243, 0.241)&(0.244, 0.239)&(0.238, 0.238)&(0.250, 0.237)\\
  &&2S Hybrid &(0.241, 0.240)&(0.239, 0.239)&(0.240, 0.239)&(0.237, 0.238)&(0.237, 0.238)&(0.233, 0.237)&(0.239, 0.236)\\
\hline
\end{tabular}
\end{center}
}
\end{table}
\end{landscape}

\pagebreak
\renewcommand{\baselinestretch}{1.1}
\begin{table}[htbp]
{\footnotesize
\caption{Simulation-based relative efficiency results for simple and optimized estimators under various one-stage (1S) and two-stage (2S) designs with $n_1=n_0=150$ (see Appendix D for details).}\label{sim.sss.re}
\newcolumntype{d}{D{.}{.}{2}}
\newcolumntype{e}{D{.}{.}{1}}
\begin{center}
\begin{tabular}{ccccddddddd}
\hline
\hline
Setting&$\gamma_1$&Design&Estimator&\multicolumn{7}{c}{$\bmX$}\\
\cline{5-11}
&&&&\multicolumn{1}{c}{$W_1$}&\multicolumn{1}{c}{$W_2$}&\multicolumn{1}{c}{$W_3$}&\multicolumn{1}{c}{$(W_1,W_2)'$}&\multicolumn{1}{c}{$(W_1,W_3)'$}&\multicolumn{1}{c}{$(W_2,W_3)'$}&\multicolumn{1}{c}{$\bmW$}\\
\hline	
 1&1&1S CIR&simple&0.94&0.95&0.94&0.94&0.95&0.95&0.95\\
      &&&optimized&1.00&1.00&1.00&1.00&1.00&1.00&1.00\\
\cline{3-11}
   &&2S CIR&simple&1.00&1.02&1.01&0.99&1.03&1.00&0.99\\
      &&&optimized&1.13&1.14&1.12&1.08&1.14&1.13&1.15\\
\cline{3-11}
   &&2S CDR&simple&1.01&1.02&1.00&0.97&1.00&0.95&0.84\\
      &&&optimized&1.11&1.13&1.13&1.08&1.13&1.15&1.10\\
\cline{2-11}
  &2&1S CIR&simple&0.95&0.95&0.95&0.96&0.95&0.95&0.95\\
      &&&optimized&1.00&1.00&1.00&1.00&1.00&1.00&1.00\\
\cline{3-11}
   &&2S CIR&simple&1.01&1.01&1.02&0.99&1.04&0.98&1.02\\
      &&&optimized&1.16&1.15&1.15&1.11&1.17&1.12&1.17\\
\cline{3-11}
   &&2S CDR&simple&1.06&1.04&1.02&0.99&1.01&0.90&0.83\\
      &&&optimized&1.16&1.17&1.15&1.12&1.15&1.09&1.07\\
\hline	
 2&1&1S CIR&simple&0.90&0.88&0.87&0.88&0.89&0.85&0.85\\
      &&&optimized&1.00&1.00&1.00&1.00&1.00&1.00&1.00\\
\cline{3-11}
   &&1S CDR&simple&0.86&0.83&0.81&0.81&0.79&0.67&0.47\\
      &&&optimized&0.87&0.57&0.77&0.87&0.90&0.58&0.52\\
\cline{3-11}
   &&2S CIR&simple&0.91&0.91&0.87&0.92&0.90&0.86&0.85\\
      &&&optimized&1.05&1.07&1.02&1.06&1.06&1.01&1.04\\
\cline{3-11}
   &&2S CDR&simple&0.87&0.87&0.86&0.91&0.84&0.74&0.60\\
      &&&optimized&1.01&1.03&1.03&1.09&1.04&0.99&0.72\\
\cline{3-11}
&&2S Hybrid&simple&0.89&0.91&0.87&0.88&0.90&0.85&0.75\\
      &&&optimized&1.04&1.07&1.02&1.09&1.12&1.08&1.05\\
\cline{2-11}
  &2&1S CIR&simple&0.90&0.89&0.87&0.86&0.85&0.86&0.85\\
      &&&optimized&1.00&1.00&1.00&1.00&1.00&1.00&1.00\\
\cline{3-11}
   &&1S CDR&simple&0.88&0.86&0.82&0.72&0.78&0.63&0.49\\
      &&&optimized&0.95&0.74&0.93&0.46&0.48&0.80&0.60\\
\cline{3-11}
   &&2S CIR&simple&0.91&0.94&0.89&0.86&0.86&0.87&0.85\\
      &&&optimized&1.05&1.09&1.05&1.03&1.00&1.05&1.03\\
\cline{3-11}
   &&2S CDR&simple&0.90&0.90&0.87&0.80&0.80&0.70&0.59\\
      &&&optimized&1.05&1.08&1.04&1.01&1.00&0.95&0.93\\
\cline{3-11}
&&2S Hybrid&simple&0.96&0.89&0.91&0.86&0.86&0.80&0.76\\
      &&&optimized&1.10&1.05&1.06&1.02&1.05&1.05&1.06\\
\hline
\end{tabular}
\end{center}
}
\end{table}

\pagebreak
\renewcommand{\baselinestretch}{1.1}
\begin{table}[htbp]
{\footnotesize
\caption{Simulation-based relative efficiency for simple and optimized estimators under one-stage (1S) and two-stage (2S) designs with various stage 1 sample size proportions ($n_1/n$) in Setting 1 (see Appendix D for details).}\label{sim.vary.samsize}
\newcolumntype{d}{D{.}{.}{2}}
\newcolumntype{e}{D{.}{.}{1}}
\begin{center}
\begin{tabular}{ccccddddddd}
\hline
\hline
$n_1/n$&$\gamma_1$&Design&Estimator&\multicolumn{7}{c}{$\bmX$}\\
\cline{5-11}
&&&&\multicolumn{1}{c}{$W_1$}&\multicolumn{1}{c}{$W_2$}&\multicolumn{1}{c}{$W_3$}&\multicolumn{1}{c}{$(W_1,W_2)'$}&\multicolumn{1}{c}{$(W_1,W_3)'$}&\multicolumn{1}{c}{$(W_2,W_3)'$}&\multicolumn{1}{c}{$\bmW$}\\
\hline
0.3&1&1S CIR&simple&0.95&0.95&0.95&0.95&0.94&0.95&0.94\\
       &&&optimized&1.00&1.00&1.00&1.00&1.00&1.00&1.00\\
\cline{3-11}
    &&2S CIR&simple&1.03&1.04&0.96&1.01&1.00&0.98&0.96\\
       &&&optimized&1.14&1.16&1.08&1.14&1.14&1.12&1.11\\
\cline{3-11}
    &&2S CDR&simple&0.99&0.98&0.96&0.95&0.96&0.90&0.69\\
       &&&optimized&1.11&1.12&1.10&1.12&1.10&1.14&1.04\\
\cline{2-11}
   &2&1S CIR&simple&0.96&0.95&0.95&0.96&0.94&0.94&0.94\\
       &&&optimized&1.00&1.00&1.00&1.00&1.00&1.00&1.00\\
\cline{3-11}
    &&2S CIR&simple&1.00&1.02&1.02&1.01&0.99&1.02&1.05\\
       &&&optimized&1.12&1.13&1.15&1.15&1.13&1.16&1.23\\
\cline{3-11}
    &&2S CDR&simple&0.98&0.98&1.01&0.94&0.93&0.84&0.76\\
       &&&optimized&1.07&1.12&1.14&1.08&1.09&1.09&1.08\\
\hline
0.7&1&1S CIR&simple&0.95&0.95&0.95&0.95&0.95&0.95&0.95\\
       &&&optimized&1.00&1.00&1.00&1.00&1.00&1.00&1.00\\
\cline{3-11}
    &&2S CIR&simple&1.00&0.99&0.98&1.00&0.99&0.99&0.99\\
       &&&optimized&1.11&1.11&1.09&1.11&1.11&1.11&1.11\\
\cline{3-11}
    &&2S CDR&simple&0.99&0.99&0.99&0.98&0.98&0.96&0.93\\
       &&&optimized&1.11&1.10&1.10&1.12&1.11&1.11&1.12\\
\cline{2-11}
   &2&1S CIR&simple&0.94&0.94&0.94&0.95&0.95&0.95&0.95\\
       &&&optimized&1.00&1.00&1.00&1.00&1.00&1.00&1.00\\
\cline{3-11}
    &&2S CIR&simple&1.01&1.01&1.01&0.99&0.99&0.98&0.98\\
       &&&optimized&1.11&1.11&1.11&1.10&1.10&1.11&1.11\\
\cline{3-11}
    &&2S CDR&simple&1.00&1.01&1.01&0.98&0.98&0.95&0.92\\
       &&&optimized&1.10&1.11&1.11&1.10&1.11&1.10&1.12\\
\hline
\end{tabular}
\end{center}
}
\end{table}

\pagebreak
\renewcommand{\baselinestretch}{1.1}
\begin{table}[htbp]
{\footnotesize
\caption{Simulation-based relative efficiency for simple and optimized estimators under one-stage (1S) and two-stage (2S) designs with various numbers of prognostic covariates (\#prog) (see Appendix D for details).}\label{sim.nonprogcov}
\newcolumntype{d}{D{.}{.}{2}}
\newcolumntype{e}{D{.}{.}{1}}
\begin{center}
\begin{tabular}{ccccddddddd}
\hline
\hline
\#prog&$\gamma_1$&Design&Estimator&\multicolumn{7}{c}{$\bmX$}\\
\cline{5-11}
&&&&\multicolumn{1}{c}{$W_1$}&\multicolumn{1}{c}{$W_2$}&\multicolumn{1}{c}{$W_3$}&\multicolumn{1}{c}{$(W_1,W_2)'$}&\multicolumn{1}{c}{$(W_1,W_3)'$}&\multicolumn{1}{c}{$(W_2,W_3)'$}&\multicolumn{1}{c}{$\bmW$}\\
\hline	
  1&1&1S CIR&simple&0.97&0.97&0.97&0.97&0.97&0.97&0.97\\
       &&&optimized&1.00&1.00&1.00&1.00&1.00&1.00&1.00\\
\cline{3-11}
    &&2S CIR&simple&1.01&1.01&1.00&1.01&1.00&1.00&1.00\\
       &&&optimized&1.08&1.09&1.09&1.09&1.09&1.09&1.09\\
\cline{3-11}
    &&2S CDR&simple&1.00&1.01&0.99&1.00&0.99&0.99&0.98\\
       &&&optimized&1.08&1.09&1.12&1.08&1.11&1.11&1.11\\
\cline{2-11}
   &2&1S CIR&simple&0.98&0.98&0.98&0.98&0.98&0.98&0.98\\
       &&&optimized&1.00&1.00&1.00&1.00&1.00&1.00&1.00\\
\cline{3-11}
    &&2S CIR&simple&0.99&0.99&0.99&0.99&0.99&1.00&1.00\\
       &&&optimized&1.06&1.06&1.08&1.07&1.08&1.08&1.08\\
\cline{3-11}
    &&2S CDR&simple&0.99&1.00&0.98&0.99&0.97&0.98&0.96\\
       &&&optimized&1.06&1.08&1.11&1.06&1.10&1.11&1.09\\
\hline
  2&1&1S CIR&simple&0.94&0.94&0.94&0.94&0.94&0.94&0.94\\
       &&&optimized&1.00&1.00&1.00&1.00&1.00&1.00&1.00\\
\cline{3-11}
    &&2S CIR&simple&0.97&0.96&0.96&0.96&0.96&0.94&0.95\\
       &&&optimized&1.08&1.09&1.09&1.09&1.08&1.08&1.09\\
\cline{3-11}
    &&2S CDR&simple&0.98&0.98&0.98&0.97&0.97&0.89&0.88\\
       &&&optimized&1.09&1.12&1.13&1.10&1.11&1.14&1.12\\
\cline{2-11}
   &2&1S CIR&simple&0.94&0.94&0.94&0.94&0.94&0.94&0.94\\
       &&&optimized&1.00&1.00&1.00&1.00&1.00&1.00&1.00\\
\cline{3-11}
    &&2S CIR&simple&0.97&0.97&0.96&0.97&0.97&0.95&0.95\\
       &&&optimized&1.08&1.09&1.08&1.08&1.08&1.08&1.08\\
\cline{3-11}
    &&2S CDR&simple&0.97&0.98&0.97&0.97&0.96&0.90&0.90\\
       &&&optimized&1.10&1.12&1.12&1.10&1.10&1.14&1.13\\
\hline
\end{tabular}
\end{center}
}
\end{table}

\pagebreak
\renewcommand{\baselinestretch}{1.1}
\begin{table}[htbp]
{\footnotesize
\caption{Simulation-based relative efficiency results for simple and optimized estimators under various one-stage (1S) and two-stage (2S) designs, under a severely misspecified outcome mean model (see Appendix D for details).}
\newcolumntype{d}{D{.}{.}{2}}
\newcolumntype{e}{D{.}{.}{1}}
\begin{center}
\begin{tabular}{ccccddddddd}
\hline
\hline
Setting&$\gamma_1$&Design&Estimator&\multicolumn{7}{c}{$\bmX$}\\
\cline{5-11}
&&&&\multicolumn{1}{c}{$W_1$}&\multicolumn{1}{c}{$W_2$}&\multicolumn{1}{c}{$W_3$}&\multicolumn{1}{c}{$(W_1,W_2)'$}&\multicolumn{1}{c}{$(W_1,W_3)'$}&\multicolumn{1}{c}{$(W_2,W_3)'$}&\multicolumn{1}{c}{$\bmW$}\\
\hline
 1&1&1S CIR&simple&0.95&0.95&0.95&0.95&0.95&0.95&0.95\\
      &&&optimized&1.00&1.00&1.00&1.00&1.00&1.00&1.00\\
\cline{3-11}
   &&2S CIR&simple&0.99&0.99&0.98&0.99&0.98&0.99&0.99\\
      &&&optimized&1.12&1.12&1.11&1.12&1.11&1.12&1.12\\
\cline{3-11}
   &&2S CDR&simple&1.03&1.03&1.02&1.02&1.01&0.99&0.98\\
      &&&optimized&1.13&1.14&1.13&1.13&1.12&1.11&1.10\\
\cline{2-11}
  &2&1S CIR&simple&0.95&0.95&0.95&0.95&0.95&0.95&0.95\\
      &&&optimized&1.00&1.00&1.00&1.00&1.00&1.00&1.00\\
\cline{3-11}
   &&2S CIR&simple&0.99&0.99&0.99&1.00&0.99&1.00&0.99\\
      &&&optimized&1.12&1.13&1.13&1.13&1.12&1.13&1.13\\
\cline{3-11}
   &&2S CDR&simple&1.02&1.02&1.02&1.02&1.00&0.99&0.97\\
      &&&optimized&1.14&1.14&1.13&1.14&1.12&1.12&1.10\\
\hline
\end{tabular}
\end{center}
}
\end{table}

\end{document}